\documentclass[reprint,amsmath,amssymb,aps,floatfix,superscriptaddress,showpacs,prl]{revtex4-1}

\usepackage{graphicx}
\usepackage{dcolumn}
\usepackage{bm}
\usepackage{color}

\begin{document}

\title{Direct Measurement of the Cosmic-Ray Proton Spectrum\\ 
 from 50~GeV to 10~TeV with the Calorimetric Electron Telescope\\
 on the International Space Station}

\author{O.~Adriani}
\affiliation{Department of Physics, University of Florence, Via Sansone, 1 - 50019 Sesto, Fiorentino, Italy}
\affiliation{INFN Sezione di Florence, Via Sansone, 1 - 50019 Sesto, Fiorentino, Italy}
\author{Y.~Akaike}
\affiliation{Department of Physics, University of Maryland, Baltimore County, 1000 Hilltop Circle, Baltimore, Maryland 21250, USA}
\affiliation{Astroparticle Physics Laboratory, NASA/GSFC, Greenbelt, Maryland 20771, USA}
\author{K.~Asano}
\affiliation{Institute for Cosmic Ray Research, The University of Tokyo, 5-1-5 Kashiwa-no-Ha, Kashiwa, Chiba 277-8582, Japan}
\author{Y.~Asaoka}
\email[]{yoichi.asaoka@aoni.waseda.jp}
\affiliation{Waseda Research Institute for Science and Engineering, Waseda University, 3-4-1 Okubo, Shinjuku, Tokyo 169-8555, Japan}
\affiliation{JEM Utilization Center, Human Spaceflight Technology Directorate, Japan Aerospace Exploration Agency, 2-1-1 Sengen, Tsukuba, Ibaraki 305-8505, Japan}
\author{M.G.~Bagliesi}
\affiliation{Department of Physical Sciences, Earth and Environment, University of Siena, via Roma 56, 53100 Siena, Italy}
\affiliation{INFN Sezione di Pisa, Polo Fibonacci, Largo B. Pontecorvo, 3 - 56127 Pisa, Italy}
\author{E.~Berti} 
\affiliation{Department of Physics, University of Florence, Via Sansone, 1 - 50019 Sesto, Fiorentino, Italy}
\affiliation{INFN Sezione di Florence, Via Sansone, 1 - 50019 Sesto, Fiorentino, Italy}
\author{G.~Bigongiari}
\affiliation{Department of Physical Sciences, Earth and Environment, University of Siena, via Roma 56, 53100 Siena, Italy}
\affiliation{INFN Sezione di Pisa, Polo Fibonacci, Largo B. Pontecorvo, 3 - 56127 Pisa, Italy}
\author{W.R.~Binns}
\affiliation{Department of Physics and McDonnell Center for the Space Sciences, Washington University, One Brookings Drive, St. Louis, Missouri 63130-4899, USA} 
\author{S.~Bonechi}
\affiliation{Department of Physical Sciences, Earth and Environment, University of Siena, via Roma 56, 53100 Siena, Italy}
\affiliation{INFN Sezione di Pisa, Polo Fibonacci, Largo B. Pontecorvo, 3 - 56127 Pisa, Italy}
\author{M.~Bongi}
\affiliation{Department of Physics, University of Florence, Via Sansone, 1 - 50019 Sesto, Fiorentino, Italy}
\affiliation{INFN Sezione di Florence, Via Sansone, 1 - 50019 Sesto, Fiorentino, Italy}
\author{A.~Bruno} 
\affiliation{Heliospheric Physics Laboratory, NASA/GSFC, Greenbelt, Maryland 20771, USA} 
\author{J.H.~Buckley}
\affiliation{Department of Physics and McDonnell Center for the Space Sciences, Washington University, One Brookings Drive, St. Louis, Missouri 63130-4899, USA} 
\author{N.~Cannady}
\affiliation{Department of Physics and Astronomy, Louisiana State University, 202 Nicholson Hall, Baton Rouge, Louisiana 70803, USA}
\author{G.~Castellini}
\affiliation{Institute of Applied Physics (IFAC),  National Research Council (CNR), Via Madonna del Piano, 10, 50019 Sesto, Fiorentino, Italy}
\author{C.~Checchia}
\affiliation{Department of Physics and Astronomy, University of Padova, Via Marzolo, 8, 35131 Padova, Italy}
\affiliation{INFN Sezione di Padova, Via Marzolo, 8, 35131 Padova, Italy} 
\author{M.L.~Cherry}
\affiliation{Department of Physics and Astronomy, Louisiana State University, 202 Nicholson Hall, Baton Rouge, Louisiana 70803, USA}
\author{G.~Collazuol}
\affiliation{Department of Physics and Astronomy, University of Padova, Via Marzolo, 8, 35131 Padova, Italy}
\affiliation{INFN Sezione di Padova, Via Marzolo, 8, 35131 Padova, Italy} 
\author{V.~Di~Felice}
\affiliation{University of Rome ``Tor Vergata'', Via della Ricerca Scientifica 1, 00133 Rome, Italy}
\affiliation{INFN Sezione di Rome ``Tor Vergata'', Via della Ricerca Scientifica 1, 00133 Rome, Italy}
\author{K.~Ebisawa}
\affiliation{Institute of Space and Astronautical Science, Japan Aerospace Exploration Agency, 3-1-1 Yoshinodai, Chuo, Sagamihara, Kanagawa 252-5210, Japan}
\author{H.~Fuke}
\affiliation{Institute of Space and Astronautical Science, Japan Aerospace Exploration Agency, 3-1-1 Yoshinodai, Chuo, Sagamihara, Kanagawa 252-5210, Japan}
\author{T.G.~Guzik}
\affiliation{Department of Physics and Astronomy, Louisiana State University, 202 Nicholson Hall, Baton Rouge, Louisiana 70803, USA}
\author{T.~Hams}
\affiliation{Department of Physics, University of Maryland, Baltimore County, 1000 Hilltop Circle, Baltimore, Maryland 21250, USA}
\affiliation{CRESST and Astroparticle Physics Laboratory NASA/GSFC, Greenbelt, Maryland 20771, USA}
\author{N.~Hasebe}
\affiliation{Waseda Research Institute for Science and Engineering, Waseda University, 3-4-1 Okubo, Shinjuku, Tokyo 169-8555, Japan}
\author{K.~Hibino}
\affiliation{Kanagawa University, 3-27-1 Rokkakubashi, Kanagawa, Yokohama, Kanagawa 221-8686, Japan}
\author{M.~Ichimura}
\affiliation{Faculty of Science and Technology, Graduate School of Science and Technology, Hirosaki University, 3, Bunkyo, Hirosaki, Aomori 036-8561, Japan}
\author{K.~Ioka}
\affiliation{Yukawa Institute for Theoretical Physics, Kyoto University, Kitashirakawa Oiwakecho, Sakyo, Kyoto 606-8502, Japan}
\author{W.~Ishizaki}
\affiliation{Institute for Cosmic Ray Research, The University of Tokyo, 5-1-5 Kashiwa-no-Ha, Kashiwa, Chiba 277-8582, Japan}
\author{M.H.~Israel}
\affiliation{Department of Physics and McDonnell Center for the Space Sciences, Washington University, One Brookings Drive, St. Louis, Missouri 63130-4899, USA} 
\author{K.~Kasahara}
\affiliation{Waseda Research Institute for Science and Engineering, Waseda University, 3-4-1 Okubo, Shinjuku, Tokyo 169-8555, Japan}
\author{J.~Kataoka}
\affiliation{Waseda Research Institute for Science and Engineering, Waseda University, 3-4-1 Okubo, Shinjuku, Tokyo 169-8555, Japan}
\author{R.~Kataoka}
\affiliation{National Institute of Polar Research, 10-3, Midori-cho, Tachikawa, Tokyo 190-8518, Japan}
\author{Y.~Katayose}
\affiliation{Faculty of Engineering, Division of Intelligent Systems Engineering, Yokohama National University, 79-5 Tokiwadai, Hodogaya, Yokohama 240-8501, Japan}
\author{C.~Kato}
\affiliation{Faculty of Science, Shinshu University, 3-1-1 Asahi, Matsumoto, Nagano 390-8621, Japan}
\author{N.~Kawanaka}
\affiliation{Hakubi Center, Kyoto University, Yoshida Honmachi, Sakyo-ku, Kyoto 606-8501, Japan}
\affiliation{Department of Astronomy, Graduate School of Science, Kyoto University, Kitashirakawa Oiwake-cho, Sakyo-ku, Kyoto 606-8502, Japan}
\author{Y.~Kawakubo}
\affiliation{Department of Physics and Astronomy, Louisiana State University, 202 Nicholson Hall, Baton Rouge, Louisiana 70803, USA} 
\affiliation{College of Science and Engineering, Department of Physics and Mathematics, Aoyama Gakuin University,  5-10-1 Fuchinobe, Chuo, Sagamihara, Kanagawa 252-5258, Japan}
\author{K.~Kohri} 
\affiliation{Institute of Particle and Nuclear Studies, High Energy Accelerator Research Organization, 1-1 Oho, Tsukuba, Ibaraki 305-0801, Japan} 
\author{H.S.~Krawczynski}
\affiliation{Department of Physics and McDonnell Center for the Space Sciences, Washington University, One Brookings Drive, St. Louis, Missouri 63130-4899, USA} 
\author{J.F.~Krizmanic}
\affiliation{CRESST and Astroparticle Physics Laboratory NASA/GSFC, Greenbelt, Maryland 20771, USA}
\affiliation{Department of Physics, University of Maryland, Baltimore County, 1000 Hilltop Circle, Baltimore, Maryland 21250, USA}
\author{T.~Lomtadze}
\affiliation{INFN Sezione di Pisa, Polo Fibonacci, Largo B. Pontecorvo, 3 - 56127 Pisa, Italy}
\author{P.~Maestro}
\affiliation{Department of Physical Sciences, Earth and Environment, University of Siena, via Roma 56, 53100 Siena, Italy}
\affiliation{INFN Sezione di Pisa, Polo Fibonacci, Largo B. Pontecorvo, 3 - 56127 Pisa, Italy}
\author{P.S.~Marrocchesi}
\email[]{piersimone.marrocchesi@pi.infn.it}
\affiliation{Department of Physical Sciences, Earth and Environment, University of Siena, via Roma 56, 53100 Siena, Italy}
\affiliation{INFN Sezione di Pisa, Polo Fibonacci, Largo B. Pontecorvo, 3 - 56127 Pisa, Italy}
\author{A.M.~Messineo}
\affiliation{University of Pisa, Polo Fibonacci, Largo B. Pontecorvo, 3 - 56127 Pisa, Italy}
\affiliation{INFN Sezione di Pisa, Polo Fibonacci, Largo B. Pontecorvo, 3 - 56127 Pisa, Italy}
\author{J.W.~Mitchell}
\affiliation{Astroparticle Physics Laboratory, NASA/GSFC, Greenbelt, Maryland 20771, USA}
\author{S.~Miyake}
\affiliation{Department of Electrical and Electronic Systems Engineering, National Institute of Technology, Ibaraki College, 866 Nakane, Hitachinaka, Ibaraki 312-8508 Japan}
\author{A.A.~Moiseev}
\affiliation{Department of Astronomy, University of Maryland, College Park, Maryland 20742, USA }
\affiliation{CRESST and Astroparticle Physics Laboratory NASA/GSFC, Greenbelt, Maryland 20771, USA}
\author{K.~Mori}
\affiliation{Waseda Research Institute for Science and Engineering, Waseda University, 3-4-1 Okubo, Shinjuku, Tokyo 169-8555, Japan}
\affiliation{Institute of Space and Astronautical Science, Japan Aerospace Exploration Agency, 3-1-1 Yoshinodai, Chuo, Sagamihara, Kanagawa 252-5210, Japan}
\author{M.~Mori}
\affiliation{Department of Physical Sciences, College of Science and Engineering, Ritsumeikan University, Shiga 525-8577, Japan}
\author{N.~Mori}
\affiliation{INFN Sezione di Florence, Via Sansone, 1 - 50019 Sesto, Fiorentino, Italy}
\author{H.M.~Motz}
\affiliation{Faculty of Science and Engineering, Global Center for Science and Engineering, Waseda University, 3-4-1 Okubo, Shinjuku, Tokyo 169-8555, Japan} 
\author{K.~Munakata}
\affiliation{Faculty of Science, Shinshu University, 3-1-1 Asahi, Matsumoto, Nagano 390-8621, Japan}
\author{H.~Murakami}
\affiliation{Waseda Research Institute for Science and Engineering, Waseda University, 3-4-1 Okubo, Shinjuku, Tokyo 169-8555, Japan}
\author{S.~Nakahira}
\affiliation{RIKEN, 2-1 Hirosawa, Wako, Saitama 351-0198, Japan}
\author{J.~Nishimura}
\affiliation{Institute of Space and Astronautical Science, Japan Aerospace Exploration Agency, 3-1-1 Yoshinodai, Chuo, Sagamihara, Kanagawa 252-5210, Japan}
\author{G.A.~de~Nolfo}
\affiliation{Heliospheric Physics Laboratory, NASA/GSFC, Greenbelt, Maryland 20771, USA}
\author{S.~Okuno}
\affiliation{Kanagawa University, 3-27-1 Rokkakubashi, Kanagawa, Yokohama, Kanagawa 221-8686, Japan}
\author{J.F.~Ormes}
\affiliation{Department of Physics and Astronomy, University of Denver, Physics Building, Room 211, 2112 East Wesley Avenue, Denver, Colorado 80208-6900, USA}
\author{S.~Ozawa}
\affiliation{Waseda Research Institute for Science and Engineering, Waseda University, 3-4-1 Okubo, Shinjuku, Tokyo 169-8555, Japan}
\author{L.~Pacini}
\affiliation{Department of Physics, University of Florence, Via Sansone, 1 - 50019 Sesto, Fiorentino, Italy}
\affiliation{Institute of Applied Physics (IFAC),  National Research Council (CNR), Via Madonna del Piano, 10, 50019 Sesto, Fiorentino, Italy}
\affiliation{INFN Sezione di Florence, Via Sansone, 1 - 50019 Sesto, Fiorentino, Italy}
\author{F.~Palma}
\affiliation{University of Rome ``Tor Vergata'', Via della Ricerca Scientifica 1, 00133 Rome, Italy}
\affiliation{INFN Sezione di Rome ``Tor Vergata'', Via della Ricerca Scientifica 1, 00133 Rome, Italy}
\author{P.~Papini}
\affiliation{INFN Sezione di Florence, Via Sansone, 1 - 50019 Sesto, Fiorentino, Italy}
\author{A.V.~Penacchioni}
\affiliation{Department of Physical Sciences, Earth and Environment, University of Siena, via Roma 56, 53100 Siena, Italy}
\affiliation{ASI Science Data Center (ASDC), Via del Politecnico snc, 00133 Rome, Italy}
\author{B.F.~Rauch}
\affiliation{Department of Physics and McDonnell Center for the Space Sciences, Washington University, One Brookings Drive, St. Louis, Missouri 63130-4899, USA} 
\author{S.B.~Ricciarini}
\affiliation{Institute of Applied Physics (IFAC),  National Research Council (CNR), Via Madonna del Piano, 10, 50019 Sesto, Fiorentino, Italy}
\affiliation{INFN Sezione di Florence, Via Sansone, 1 - 50019 Sesto, Fiorentino, Italy}
\author{K.~Sakai}
\affiliation{CRESST and Astroparticle Physics Laboratory NASA/GSFC, Greenbelt, Maryland 20771, USA}
\affiliation{Department of Physics, University of Maryland, Baltimore County, 1000 Hilltop Circle, Baltimore, Maryland 21250, USA}
\author{T.~Sakamoto}
\affiliation{College of Science and Engineering, Department of Physics and Mathematics, Aoyama Gakuin University,  5-10-1 Fuchinobe, Chuo, Sagamihara, Kanagawa 252-5258, Japan}
\author{M.~Sasaki}
\affiliation{CRESST and Astroparticle Physics Laboratory NASA/GSFC, Greenbelt, Maryland 20771, USA}
\affiliation{Department of Astronomy, University of Maryland, College Park, Maryland 20742, USA }
\author{Y.~Shimizu}
\affiliation{Kanagawa University, 3-27-1 Rokkakubashi, Kanagawa, Yokohama, Kanagawa 221-8686, Japan}
\author{A.~Shiomi}
\affiliation{College of Industrial Technology, Nihon University, 1-2-1 Izumi, Narashino, Chiba 275-8575, Japan}
\author{R.~Sparvoli}
\affiliation{University of Rome ``Tor Vergata'', Via della Ricerca Scientifica 1, 00133 Rome, Italy}
\affiliation{INFN Sezione di Rome ``Tor Vergata'', Via della Ricerca Scientifica 1, 00133 Rome, Italy}
\author{P.~Spillantini}
\affiliation{Department of Physics, University of Florence, Via Sansone, 1 - 50019 Sesto, Fiorentino, Italy}
\author{F.~Stolzi}
\affiliation{Department of Physical Sciences, Earth and Environment, University of Siena, via Roma 56, 53100 Siena, Italy}
\affiliation{INFN Sezione di Pisa, Polo Fibonacci, Largo B. Pontecorvo, 3 - 56127 Pisa, Italy}
\author{J.E.~Suh} 
\affiliation{Department of Physical Sciences, Earth and Environment, University of Siena, via Roma 56, 53100 Siena, Italy}
\affiliation{INFN Sezione di Pisa, Polo Fibonacci, Largo B. Pontecorvo, 3 - 56127 Pisa, Italy}
\author{A.~Sulaj} 
\affiliation{Department of Physical Sciences, Earth and Environment, University of Siena, via Roma 56, 53100 Siena, Italy}
\affiliation{INFN Sezione di Pisa, Polo Fibonacci, Largo B. Pontecorvo, 3 - 56127 Pisa, Italy}
\author{I.~Takahashi}
\affiliation{Kavli Institute for the Physics and Mathematics of the Universe, The University of Tokyo, 5-1-5 Kashiwanoha, Kashiwa, 277-8583, Japan}
\author{M.~Takayanagi}
\affiliation{Institute of Space and Astronautical Science, Japan Aerospace Exploration Agency, 3-1-1 Yoshinodai, Chuo, Sagamihara, Kanagawa 252-5210, Japan}
\author{M.~Takita}
\affiliation{Institute for Cosmic Ray Research, The University of Tokyo, 5-1-5 Kashiwa-no-Ha, Kashiwa, Chiba 277-8582, Japan}
\author{T.~Tamura}
\affiliation{Kanagawa University, 3-27-1 Rokkakubashi, Kanagawa, Yokohama, Kanagawa 221-8686, Japan}
\author{T.~Terasawa}
\affiliation{RIKEN, 2-1 Hirosawa, Wako, Saitama 351-0198, Japan}
\author{H.~Tomida}
\affiliation{Institute of Space and Astronautical Science, Japan Aerospace Exploration Agency, 3-1-1 Yoshinodai, Chuo, Sagamihara, Kanagawa 252-5210, Japan}
\author{S.~Torii}
\email[]{torii.shoji@waseda.jp}
\affiliation{Waseda Research Institute for Science and Engineering, Waseda University, 3-4-1 Okubo, Shinjuku, Tokyo 169-8555, Japan}
\affiliation{School of Advanced Science and Engineering, Waseda University, 3-4-1 Okubo, Shinjuku, Tokyo 169-8555, Japan}
\author{Y.~Tsunesada}
\affiliation{Division of Mathematics and Physics, Graduate School of Science, Osaka City University, 3-3-138 Sugimoto, Sumiyoshi, Osaka 558-8585, Japan}
\author{Y.~Uchihori}
\affiliation{National Institutes for Quantum and Radiation Science and Technology, 4-9-1 Anagawa, Inage, Chiba 263-8555, Japan}
\author{S.~Ueno}
\affiliation{Institute of Space and Astronautical Science, Japan Aerospace Exploration Agency, 3-1-1 Yoshinodai, Chuo, Sagamihara, Kanagawa 252-5210, Japan}
\author{E.~Vannuccini}
\affiliation{INFN Sezione di Florence, Via Sansone, 1 - 50019 Sesto, Fiorentino, Italy}
\author{J.P.~Wefel}
\affiliation{Department of Physics and Astronomy, Louisiana State University, 202 Nicholson Hall, Baton Rouge, Louisiana 70803, USA}
\author{K.~Yamaoka}
\affiliation{Nagoya University, Furo, Chikusa, Nagoya 464-8601, Japan}
\author{S.~Yanagita}
\affiliation{College of Science, Ibaraki University, 2-1-1 Bunkyo, Mito, Ibaraki 310-8512, Japan}
\author{A.~Yoshida}
\affiliation{College of Science and Engineering, Department of Physics and Mathematics, Aoyama Gakuin University,  5-10-1 Fuchinobe, Chuo, Sagamihara, Kanagawa 252-5258, Japan}
\author{K.~Yoshida}
\affiliation{Department of Electronic Information Systems, Shibaura Institute of Technology, 307 Fukasaku, Minuma, Saitama 337-8570, Japan}

\collaboration{CALET Collaboration}

\date{\today}

\begin{abstract}
In this paper, we present the analysis and results 
of a direct measurement of the cosmic-ray proton spectrum with the CALET
instrument onboard the International Space Station,
including the detailed assessment of systematic uncertainties.
The observation period used in this analysis 
is from October 13, 2015 to August 31, 2018 (1054 days).
We have achieved the very wide energy range necessary to carry out
measurements of the spectrum from 50~GeV to 10~TeV
covering, for the first time
in space, 
with a single instrument the whole
energy interval previously investigated 
in most cases
in separate subranges by 
magnetic spectrometers 
(BESS-TeV, PAMELA, and AMS-02) 
and calorimetric instruments
(ATIC, CREAM, and NUCLEON). 
The observed spectrum is consistent with AMS-02 but extends to nearly 
an order of magnitude higher energy, showing a very smooth transition of
the power-law spectral index
from $-2.81 \pm 0.03$ 
(50--500~GeV) neglecting solar modulation effects (or $-2.87 \pm 0.06$
including solar modulation effects in the lower energy region) 
to 
$-2.56 \pm 0.04$ 
(1--10~TeV), 
thereby 
confirming the existence of spectral hardening and 
providing evidence of a deviation from a single power law by more than 3$\sigma$. 
\end{abstract}


\maketitle

\section{Introduction} 
Direct measurements of the high-energy spectra of 
each species of cosmic-ray nuclei up to the PeV
energy scale 
provide 
detailed insight into the general 
phenomenology 
of cosmic-ray acceleration and propagation
in the Galaxy.
A 
possible
charge-dependent cutoff in the nuclei spectra is hypothesized to explain the ``knee'' in the all-particle spectrum. This hypothesis 
can 
be tested directly with measurements by long duration 
space 
experiments with sufficient exposure 
and with the capability of identifying individual elements 
based on charge measurements. 

Furthermore, the spectral hardening observed in 
the 
spectra of various nuclei~\cite{ATIC2-p,CREAM-nuclei,CREAM-hardening,CREAM-I,PAMELA,AMS-02-proton,AMS-02-helium,CREAM-III-pHe,AMS-02-carbon,AMS-02-boron,AMS-02-nitrogen} calls for the extensive attempts~\cite{IP1-Ellison-1997, IP2-Malkov-2012, LS1-Erlykin-2012, LS2-Thoudam-2012, LS3-Bernard-2013, PR1-Blasi-2012, PR2-Aloisio-2013, RA1-Ptuskin-2011, RA2-Thoudam, SB1-Drury-2011, SB2-Ohira-2011, SB3-Ohira-2016, SS1-Biermann-2010, SS2-Ptuskin-2013, SS3-Zatsepin-2006, Tomassetti-2012, Vladmirov-2012, Tomassetti2015, Semikoz2018, Evoli2018, kawanaka-2018} to theoretically interpret 
these 
unexpected phenomena.
The current experimental approaches to direct measurements of the proton spectrum are 
based on 
two main classes of 
instruments, i.e., 
magnetic 
spectrometers~\cite{PAMELA,AMS-02-proton} 
at 
lower energies 
where the presence of a spectral breakpoint was observed, and 
calorimeters~\cite{ATIC2-p,CREAM-I,CREAM-III-pHe,NUCLEON-JCAP,NUCLEON-JTEP} 
at higher energies where the spectrum undergoes a hardening. 
It is of particular interest to determine the onset of spectral hardening 
and its development in terms of 
index variation and smoothness parameter (as defined in Ref.~\cite{AMS-02-proton}). 
In order to achieve a consistent picture, measurements should be unaffected,
as much as possible, by systematic errors and a critical comparison of the 
observations from different experiments is in order.

The CALorimetric Electron Telescope (CALET)~\cite{torii2017,asaoka2018},  
a space-based instrument optimized for the measurement 
of the 
all-electron spectrum~\cite{CALET2017,CALET2018} 
and 
equipped with a fully active calorimeter, 
can measure the main 
components 
of 
cosmic rays 
including proton, light and 
heavy 
nuclei 
(up to iron and above) 
in the 
energy 
range 
up to 
$\sim$1~PeV. 
The thickness of the calorimeter corresponds to 
30 radiation length (at normal incidence) and to 
$\sim$1.3 proton interaction length.

In this Letter, we present 
a direct measurement of 
the 
cosmic-ray proton spectrum 
from $E=50$~GeV to 10~TeV with CALET
where $E$ denotes the kinetic energy of primary protons 
throughout 
this paper.
Its wide dynamic range allows
the 
study of the detailed shape of the spectrum by using a single instrument.

\section{CALET Instrument}

CALET consists of a charge detector (CHD), 
a 3 radiation-length thick imaging calorimeter (IMC) and a 27 radiation-length 
thick total absorption calorimeter (TASC), 
with 
a field of view of  $\sim$45$^\circ$ from zenith. 
A ``fiducial'' geometrical factor of $\sim$416~cm$^2$sr for particles 
penetrating CHD top to TASC bottom,
with 2~cm margins at the first and the last TASC layers 
(Acceptance A), 
and corresponding to about 40\% of the 
total acceptance~\cite{CALET2018}, 
is used in this analysis.

The CHD, which identifies the charge of the incident particle, 
is comprised of a pair of plastic scintillator hodoscopes 
arranged in two orthogonal layers. 
The IMC is a sampling calorimeter alternating thin layers of Tungsten absorber 
with layers of scintillating fibers read-out individually, also providing
an independent charge measurement via multiple $dE$/$dx$ 
samples. 
The TASC is a tightly packed lead-tungstate (PbWO$_4$) hodoscope, 
measuring the energy of showering particles in the detector.
A very large 
dynamic range of more than 6 orders of magnitude 
is covered by four different 
gain ranges~\cite{asaoka2017}.
A more complete description of the instrument is given 
in the Supplemental Material of Ref.~\cite{CALET2017}.

\begin{figure}[t!]
\begin{center}
\includegraphics[width=\linewidth]{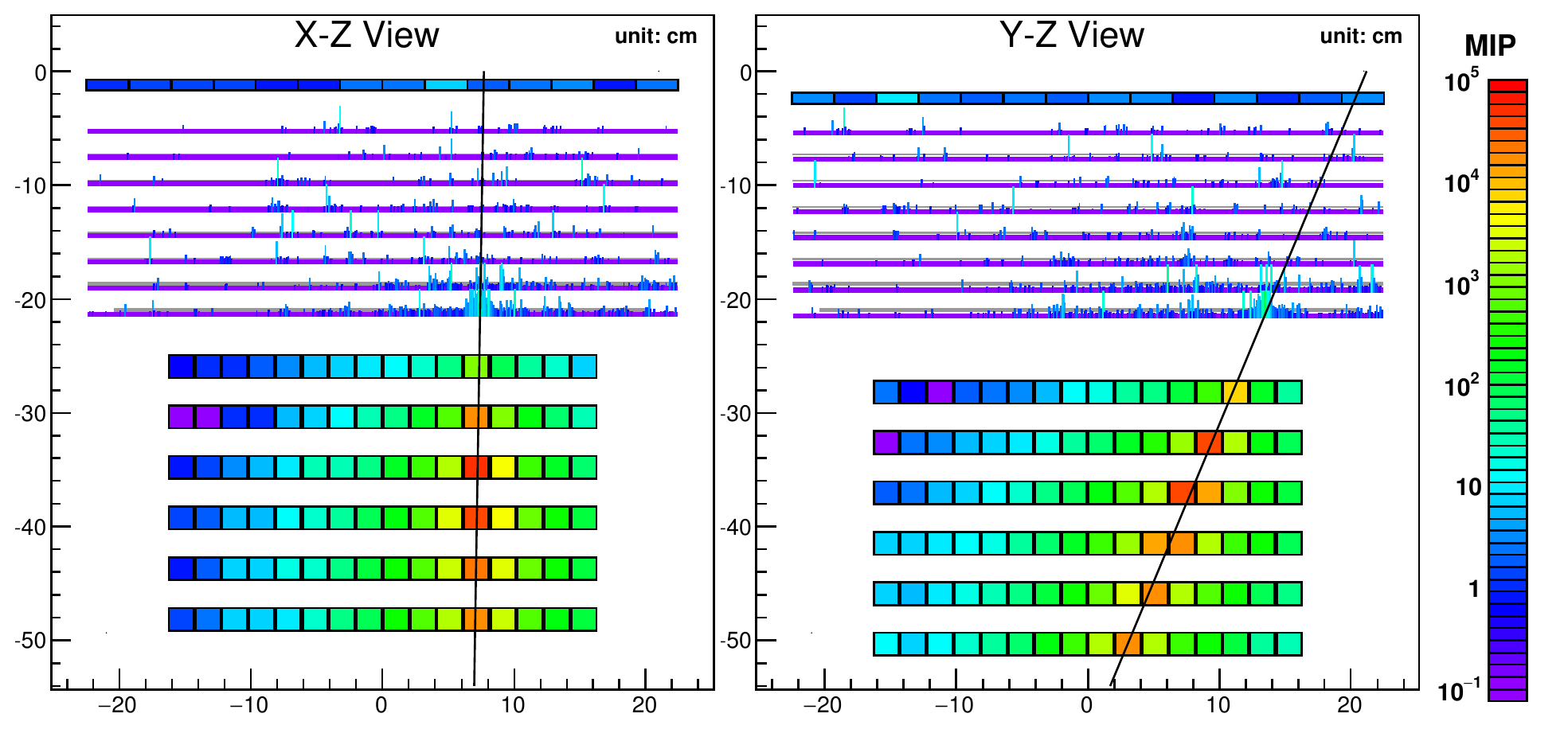}
\caption{
An example of a high-energy proton event 
with an energy deposit sum of 
10~TeV. 
Black lines represent the reconstructed tracks using 
Kalman Filter tracking~\cite{paolo2017}, which 
achieves a fine resolution taking advantage of 
the high granularity of the IMC. 
}
\label{fig:tevevt}
\end{center}
\end{figure}

Figure~\ref{fig:tevevt} shows a proton candidate with
energy deposit of $\sim$10~TeV in the detector.
The 
event example clearly demonstrates CALET's capability to reconstruct 
and identify very high energy protons.
Because of the limited energy resolution, energy unfolding is required 
to estimate the primary energy distribution.
It is important, therefore, to 
infer 
the 
detector 
response 
at the 
highest energies covered by the analysis. 

The instrument was launched on August 19, 2015 and emplaced on the 
Japanese Experiment Module-Exposed Facility on the International Space Station 
with an expected mission duration of five years (or more). 
Scientific observations~\cite{asaoka2018} 
started on October 13, 2015, 
and smooth and continuous operations have taken place since then.

\section{Data Analysis} 
We have analyzed 
flight data collected 
for 1054 days from October 13, 2015
to August 31, 2018. 
The total 
observation 
live time for high-energy (HE) shower trigger~\cite{asaoka2018}
is 21421.9 hours and live time fraction to total time is 84.7\%.
In addition, the low-energy (LE) shower trigger operated 
at a high geomagnetic latitude~\cite{asaoka2018} is used to 
extend the energy coverage toward the lower energy region.
In spite of 
a 
limited live time of 365.4~hr, 
LE data provide 
sufficient
statistics for protons below a few hundred GeV.

Monte Carlo (MC) simulations, 
reproducing the 
detailed detector configuration,
physics processes, 
as well as 
detector signals, 
are 
based on the 
EPICS simulation package \cite{EPICS,EPICSver}. 

In order to 
assess 
the relatively large uncertainties in
the hadronic interactions, a series of beam tests were carried out at CERN-SPS
using the CALET beam test model~\cite{akaike2013, niita2015, akaike2015}.
Trigger efficiency and energy response derived from MC 
simulations were tuned
using the beam test results obtained 
in 2012~\cite{akaike2013, niita2015, tamura2017} 
with proton beams of 30, 100 and 400 GeV. 
The correction for 
the 
trigger efficiency
obtained 
by the EPICS simulation 
was 
determined to be 7.7\% for the LE trigger and 11.2\% for the HE trigger,
irrespective of proton energies. 
Shower energy correction was determined to be 
7.9\% and 6.3\% 
at 
30 and 100~GeV, 
and no correction 
at 
400~GeV and above, 
where 
simple log-linear 
interpolation was used to determine the correction factor for intermediate
energies.

In the analysis of hadrons, 
especially in the high-energy region where 
no beam test calibration is possible, 
a comparison between different MC models becomes
much more important than in 
electron analysis. 
For this purpose, we have 
run simulations with
FLUKA~\cite{FLUKA-1,FLUKA-2,FLUKAver}  and Geant4~\cite{Geant4,Geant4ver} 
in the same way as EPICS.
The detector models used in FLUKA and Geant4 
are 
almost identical to
the CALET CAD model used in EPICS. 

In 
electron analysis~\cite{CALET2017,CALET2018}, 
the 
electromagnetic shower tracking algorithm works
very well, 
because of the presence of a developed shower core that is used as an initial
guess of the trajectory of the incoming particle.
In the proton analysis, however, the hadronic interaction occurs abruptly 
and there is no guarantee 
of the presence of a well-developed shower core in the bottom 
two layers of the IMC. 
It is therefore necessary to 
follow 
a different approach 
to reconstruct the proton tracks
in a highly efficient way.
Combinatorial 
Kalman Filter tracking~\cite{paolo2017} was developed for this 
purpose and 
it 
is used in the proton spectrum analysis described 
hereafter. 

The shower energy of each event is calculated as the TASC energy deposit sum 
(observed energy: $E_{\rm TASC}$), 
which is calibrated 
using penetrating particles and 
by 
performing a seamless stitching of 
adjacent gain ranges
on orbit, 
complemented by the confirmation of the linearity of the system over the whole
range by means of ground measurements using UV pulse laser
as described in Ref.~\cite{asaoka2017}.
Temporal variations 
during the long-term observation 
are 
also corrected, 
sensor by sensor, using penetrating particles as 
gain monitor~\cite{CALET2017}.

In order to minimize helium contamination 
by accurately separating protons from helium 
based on their 
charge, 
a
preselection of well-reconstructed and well-contained events 
is applied. 
Preselection consists of (1) offline trigger confirmation, 
(2) geometrical condition (requires Acceptance A~\cite{CALET2018}), 
(3) track quality cut to ensure reliability of the reconstructed track while 
retaining high efficiency, 
(4) electron rejection cut,
(5) off-acceptance events rejection cut, 
(6) requirement 
of
track consistency with TASC energy deposits,
and (7) shower development requirement in the IMC. 
Some of the above selections are described in more detail in the following.

Consistency between MC and flight data (FD)
for 
triggered events 
is 
obtained
by an offline trigger, 
which requires 
more severe conditions 
than the onboard trigger.
It removes non-negligible effects due 
to positional and temporal variation of the detector gain, and 
it is 
applied as a first step of preselection.

In order to reject electronlike events, a ``Moliere concentration'' 
along the track 
in the IMC 
is calculated by summing up all energy deposits 
found inside 
one 
Moliere radius for Tungsten 
($\pm$9 fibers, i.e., 9~mm) 
around
the fiber matched 
with
the track,
and normalized to the total
energy deposit sum in the IMC. 
By requiring 
this 
quantity to be less than 0.7, 
most of electrons are rejected 
while retaining an efficiency above 92\% for protons. 

Because of the nature of hadronic interaction and 
combinatorial 
track reconstruction, there is a possibility to 
introduce a 
misreconstruction by 
erroneously identifying 
one of the secondary tracks 
as the primary track. 
To minimize the fraction of misidentified events, 
two topological cuts are applied using TASC energy-deposit information
irrespective of IMC tracking.

Further rejection is achieved with a consistency cut 
between the track impact point and center of gravity of 
energy deposits in the first (TASC-X1) and second (TASC-Y1) 
layers of the TASC.
Energy dependent thresholds are defined using MC simulation 
to have a constant efficiency of 95\% for events 
that 
interacted 
in the IMC 
below the 
fourth layer, which are suitable 
for determining charge, 
energy, and trigger efficiency (hereafter denoted as ``target'' events).

Backscattered particles produced in the shower affect both 
the trigger and the charge determination. 
Primary particles below the trigger thresholds might be 
triggered anyway because of backscattered particles hitting 
TASC-X1 and IMC 
bottom 
layers. 
Moreover the large amount of shower particle tracks 
backscattered from TASC may induce fake charge identification 
by releasing additional amounts of energy that add up 
to the primary particle ionization signal, resulting in 
a shift of the charge distribution and a larger width. 

Since 
a fraction 
of events triggered by backscattering 
is not reproduced well by the simulations, 
rejection of such events 
is 
important.
For this purpose,
the energy deposit sum along the shower axis over $\pm$9 
fibers (in total 19 fibers) is 
used to ensure the existence of 
a 
shower core in the IMC. 
This definition differs from 
the one used for 
electrons considering the 
wider lateral spread of hadronic 
showers. 
In order to fully exploit the rejection capability of events 
triggered by backscattering, 
it is important to set an appropriate threshold as a function of energy.
Energy dependent thresholds 
are 
defined to 
get 
99\% 
efficiency
for ``target'' events.

The identification of cosmic-ray nuclei via a measurement of 
their charge is carried out 
with two independent 
subsystems 
that are routinely used to 
cross-calibrate each other: the CHD and the IMC~\cite{pier2017}. 
The latter samples the ionization deposits in each layer, thereby providing a
multiple $dE$/$dx$ measurement with a maximum of 16 samples along the track. 
The interaction point is first reconstructed~\cite{brogi2015} 
and only the $dE$/$dx$ ionization clusters from the layers upstream 
the interaction point are used. 
The charge value is evaluated as 
a truncated mean 
of the valid samples with a truncation level set at 70\%.

\begin{figure}[bt!] 
\begin{center}
\includegraphics[width=\hsize]{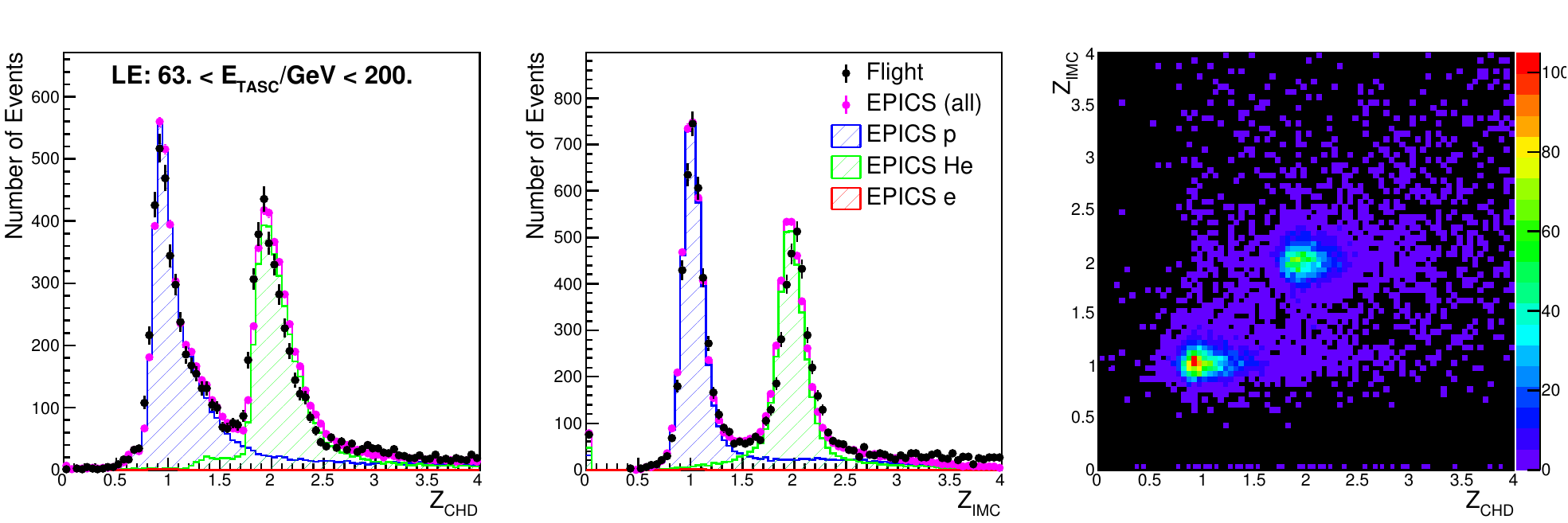}
\includegraphics[width=\hsize]{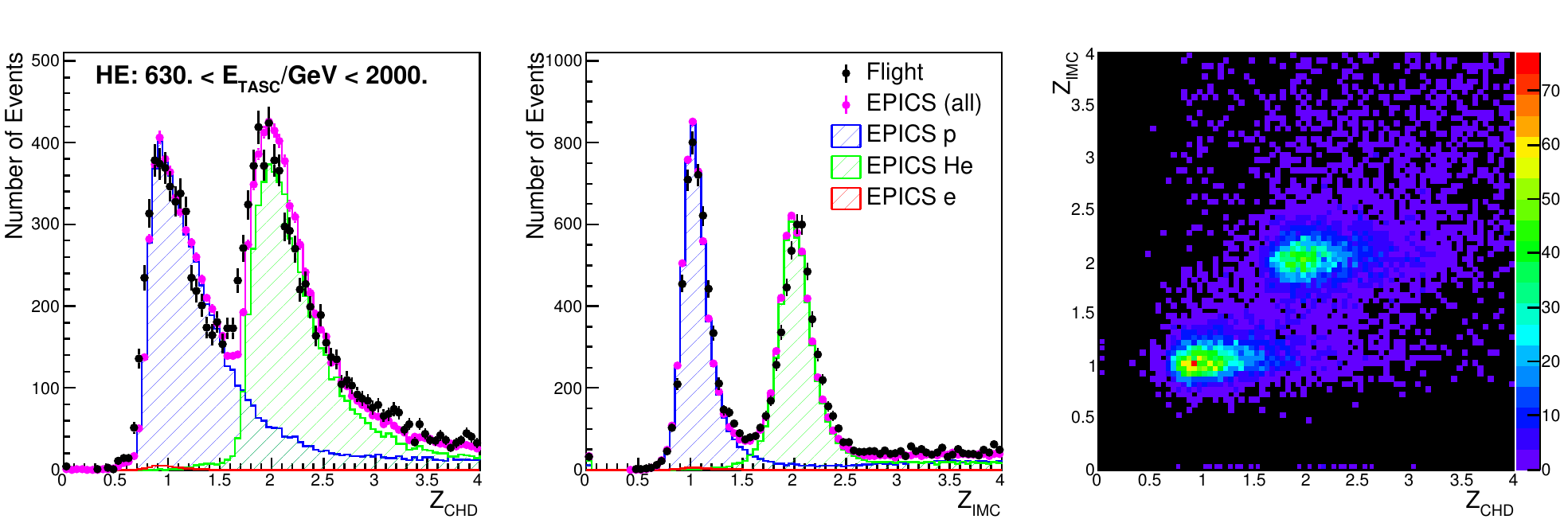}
\caption{Examples of CHD/IMC charge distributions.
Left, center, and right panels show the CHD charge, IMC charge 
and correlation between CHD and IMC charges, respectively. 
Top and bottom plots correspond to events with $63 < E_{\rm TASC} < 200$~GeV 
and $630 < E_{\rm TASC} < 2000$~GeV, respectively. 
An enlarged version of the figure is available as Fig.~S1
in the Supplemental Material~\cite{CALET2018-proton-SM}.
}
\label{fig:distZ}
\end{center}
\end{figure}

To mitigate the backscattering effects, 
an
energy dependent charge correction to 
restore the nominal peak positions
of protons and helium
to $Z=1$ and 2 is applied 
separately to 
FD, EPICS, FLUKA and Geant4,
where the same correction is used for both protons and helium. 
Charge selection of proton and helium candidates is performed 
by applying simultaneous window cuts on CHD and IMC reconstructed charges. 
The resultant charge distributions are 
exemplified
in Fig.~\ref{fig:distZ}.
For the selection 
with the CHD and IMC, energy dependent thresholds 
are 
defined 
separately 
for the CHD and IMC 
to keep 95\% efficiency for ``target'' events.

\begin{figure}[b!]
\begin{center}
\includegraphics[width=1.0 \linewidth]{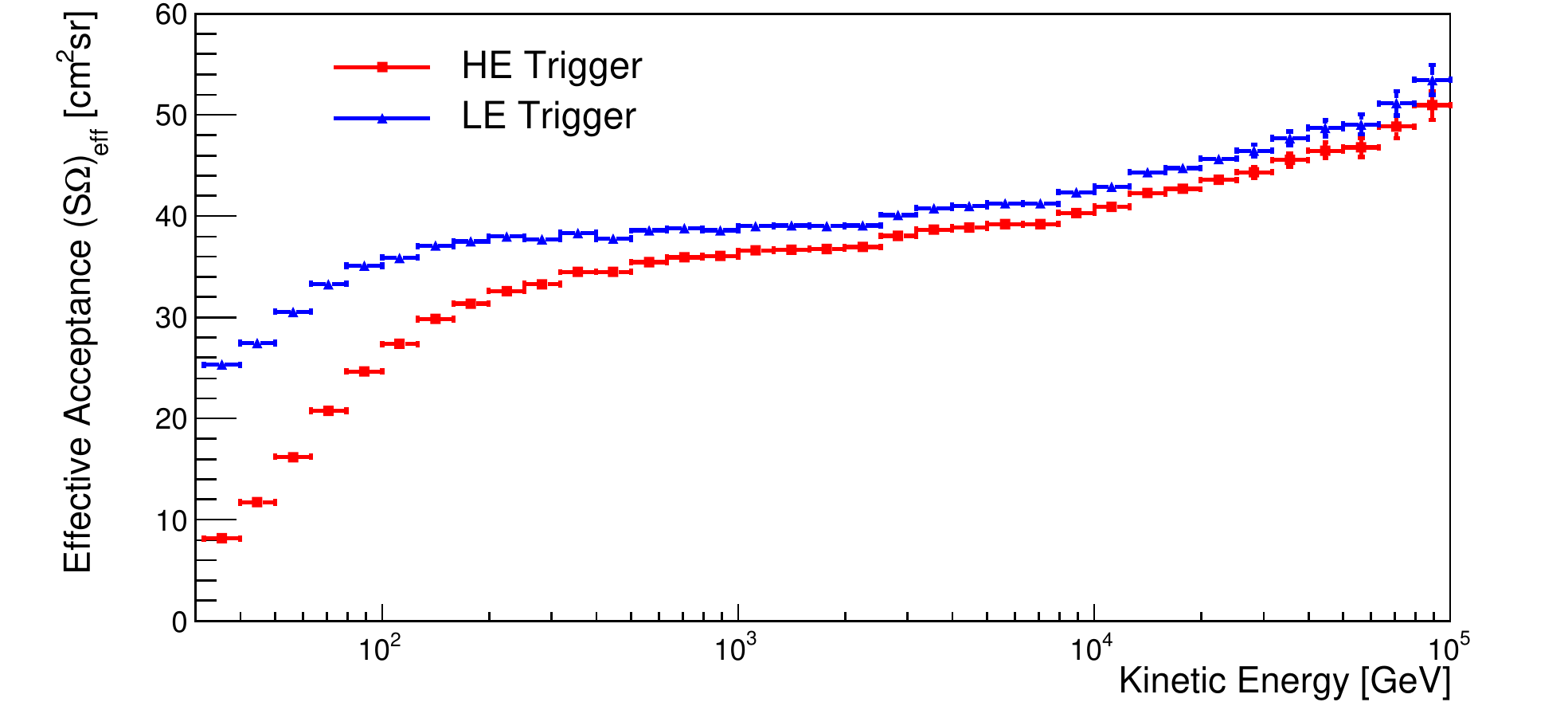}
\caption{Comparison of effective acceptance for HE-trigger (red)
and LE-trigger (blue) analyses. 
obtained by MC simulation. 
The difference between 
the two selections is the 
offline trigger confirmation only.}
\label{fig:effacc}
\end{center}
\end{figure}

In 
the 
lower energy region, the use of the LE trigger 
is necessary to avoid trigger threshold bias due to 
the 
sharp drop in efficiency
at 
$E<100$~GeV, 
an effect that extends to 
the higher energy region 
via the 
energy unfolding procedure.
With the exception of 
the offline trigger confirmation threshold 
which is 
adjusted to match
the hardware trigger, the event selection criteria 
used in  
HE and LE analyses
are 
identical.
Figure~\ref{fig:effacc} shows the effective acceptance of
LE- and HE-trigger analyses after applying all the selection criteria.
While the overall difference 
between 
the 
two analyses 
is rather small, 
the difference in the low-energy region is 
sizable. 

Background contamination 
is  estimated from the MC simulation of protons, helium, and electrons 
as a function of observed energy.
Among them, the dominant component is off-acceptance protons except for the highest
energy region $E_{\rm TASC} \sim 10$~TeV, 
where helium contamination becomes 
dominant. 
Overall contamination is estimated 
below a few percent, and at maximum 
$\sim$5\% 
in the lowest and highest energy region.
The correction is carried out before performing 
the 
energy unfolding procedure,
which is described in the following.

In order to take into account the relatively limited energy resolution
(observed energy fraction is around 35\% and the resultant energy resolution
is 30\%--40\%), 
energy unfolding is necessary to correct for bin-to-bin migration effects.
In this analysis, we used 
the Bayesian approach implemented in the 
RooUnfold 
package~\cite{roounfold, BayesUnfolding} 
in ROOT~\cite{root}, 
with the 
response matrix
derived using MC 
simulation. 
Convergence is obtained within two iterations, 
given the relatively accurate 
prior distribution obtained from 
the previous observations, i.e., AMS-02~\cite{AMS-02-proton} and 
CREAM-III~\cite{CREAM-III-pHe}.

The proton spectrum is obtained by 
correcting the effective geometrical acceptance with 
the unfolded energy distribution as follows:
\begin{eqnarray*}
 \Phi(E) & = & \frac{n(E)}{(S \Omega)_{\rm eff}(E) \,\, T \, \Delta E},\\
\label{fluxcalc}
n(E) & = & U\bigl(n_{\rm obs}(E_{\rm TASC}) - n_{\rm bg}(E_{\rm TASC})\bigr),
\label{fluxcalc2}
\end{eqnarray*}
where 
$\Delta E$ 
denotes 
the  
energy bin width,
$U()$ 
the  
unfolding procedure based on Bayes theorem,
$n(E)$
the 
bin counts of the unfolded distribution, 
$n_{\rm obs}(E_{\rm TASC})$ 
those 
of observed energy distribution (including background), 
$n_{\rm bg}(E_{\rm TASC})$ 
the 
bin 
counts 
of background events in 
the 
observed energy distribution,
$(S\Omega)_{\rm eff}$ 
the 
effective acceptance including all 
selection 
efficiencies, and
$T$ 
the 
live time.

Depending on the on-orbit trigger mode and corresponding offline-trigger threshold,
two spectra are obtained with the LE and HE analyses, respectively, 
as shown in Fig.~S2 in the Supplemental Material~\cite{CALET2018-proton-SM}. 
For $E < 200$~GeV, 
the use of LE-trigger analysis is required
because 
an offline trigger threshold higher than in the hardware trigger
was found to introduce 
an 
efficiency bias in the HE-trigger analysis, 
which became evident 
with a scan of the 
offline-trigger threshold using LE-trigger data.
Since both fluxes are well consistent 
in $E > 200$~GeV, 
they are combined 
around 
$E \sim 300$~GeV, 
taking into account the different statistics of the two trigger modes. 

\section{Systematic Uncertainties}
Dominant sources of 
systematic 
uncertainties in proton analysis include 
(1) hadronic interaction modeling, 
(2) energy response,
(3) track reconstruction, and 
(4) charge identification.
To address these uncertainties, various approaches are used 
as 
discussed 
in the Supplemental Material~\cite{CALET2018-proton-SM}. 
An important part of systematics comes from the accuracy
of the beam test calibration and its extrapolation or interpolation. 
The stability of the measured spectrum against variations of several analysis cuts
is also a crucial tool to estimate the associated 
uncertainties. 

Considering all of the above contributions, 
the total systematic uncertainty,
as 
summarized in 
Fig.~S4 in the Supplemental Material~\cite{CALET2018-proton-SM},
is within 10\% and estimated separately for
normalization and energy dependent 
uncertainties. 

\section{Results}
\begin{figure}[b!]
\begin{center}
\includegraphics[width=\hsize]{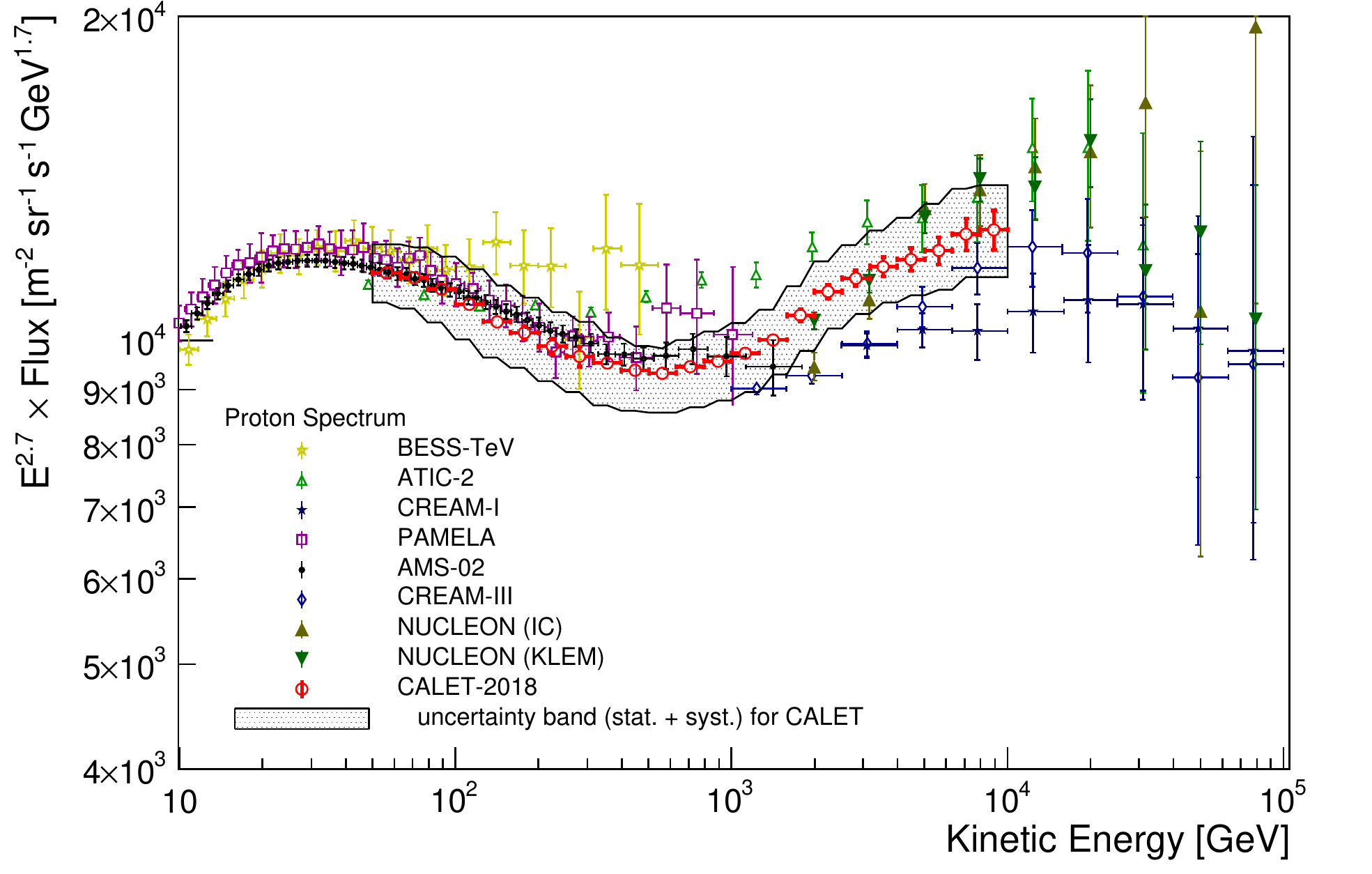}
\caption{Cosmic-ray proton spectrum measured by CALET 
(red points) 
from 50~GeV to 10~TeV.
The gray band indicates the quadratic sum of statistical and systematic errors. 
Also plotted are recent direct 
measurements~\cite{ATIC2-p,CREAM-I,AMS-02-proton,CREAM-III-pHe,BESS-TeV,PAMELA-rep,PAMELA-10yrs,NUCLEON-JTEP}.
An enlarged version of the figure is available as 
Fig.~S6
in the Supplemental Material~\cite{CALET2018-proton-SM}.
} 
\label{fig:proton}
\end{center}
\end{figure}
Figure~\ref{fig:proton} shows the proton spectrum measured with CALET 
in an energy range from 50~GeV to 10~TeV, where current 
uncertainties 
that include 
statistical and systematic errors 
are 
bounded within 
a gray band. 
The measured proton flux 
and the statistical and systematic errors are tabulated 
in Table~I 
of the Supplemental Material~\cite{CALET2018-proton-SM}. 
In Fig.~\ref{fig:proton}, the CALET spectrum is compared with recent experiments from space 
(PAMELA~\cite{PAMELA-rep,PAMELA-10yrs}, AMS-02~\cite{AMS-02-proton}, and NUCLEON~\cite{NUCLEON-JTEP}) 
and from the 
high altitude balloon experiments
(BESS-TeV~\cite{BESS-TeV}, ATIC-2~\cite{ATIC2-p}, CREAM-I~\cite{CREAM-I}, and CREAM-III~\cite{CREAM-III-pHe}). 
Our spectrum is 
in good agreement
with the 
very 
accurate 
magnetic 
spectrometer measurements by AMS-02
in the low-energy region, and 
the 
spectral behavior is also 
consistent with 
measurements from 
calorimetric
instruments in the higher energy region.

\begin{figure}[bth!]
\begin{center}
\includegraphics[width=\hsize]{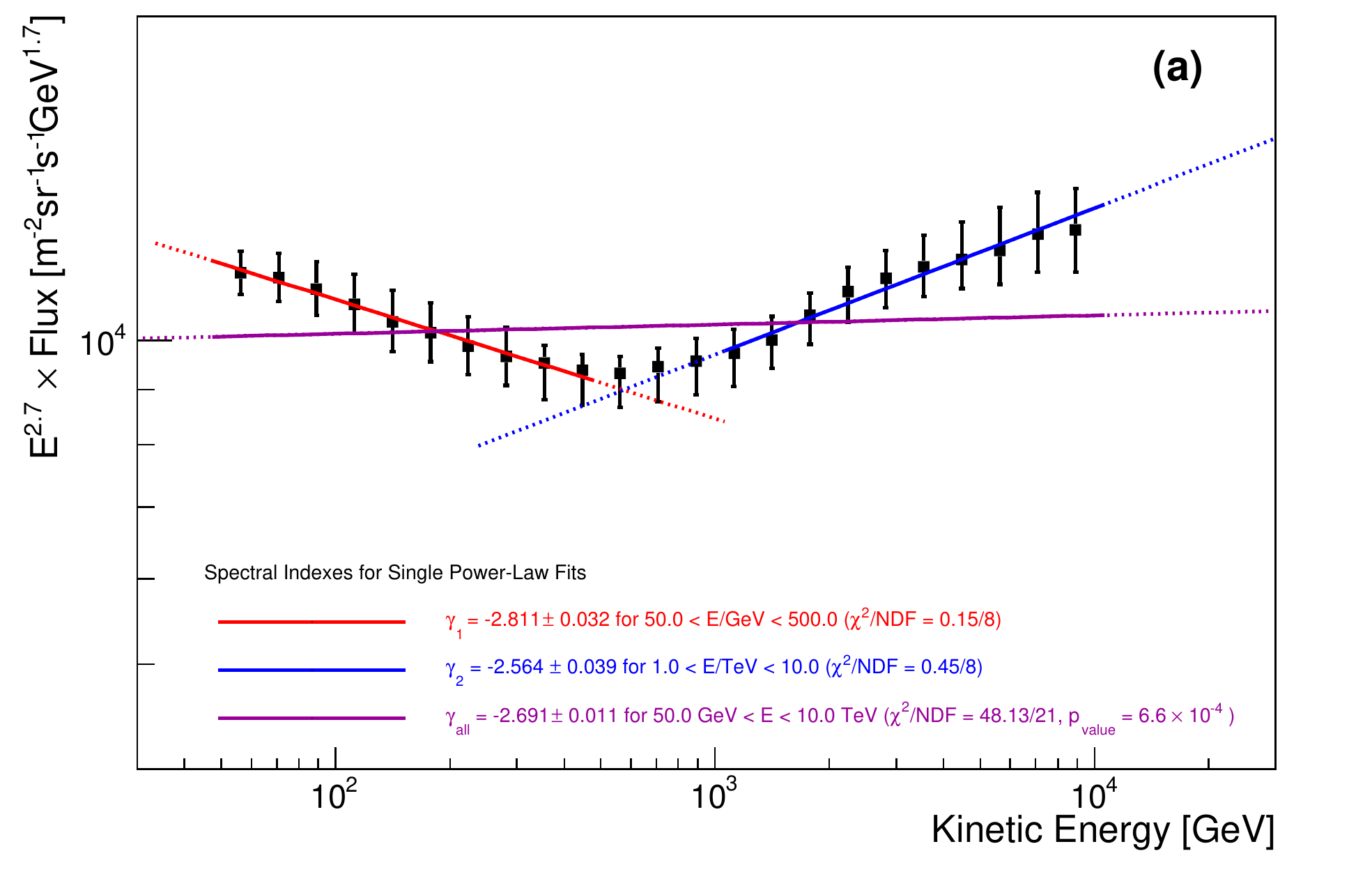}
\includegraphics[width=\hsize]{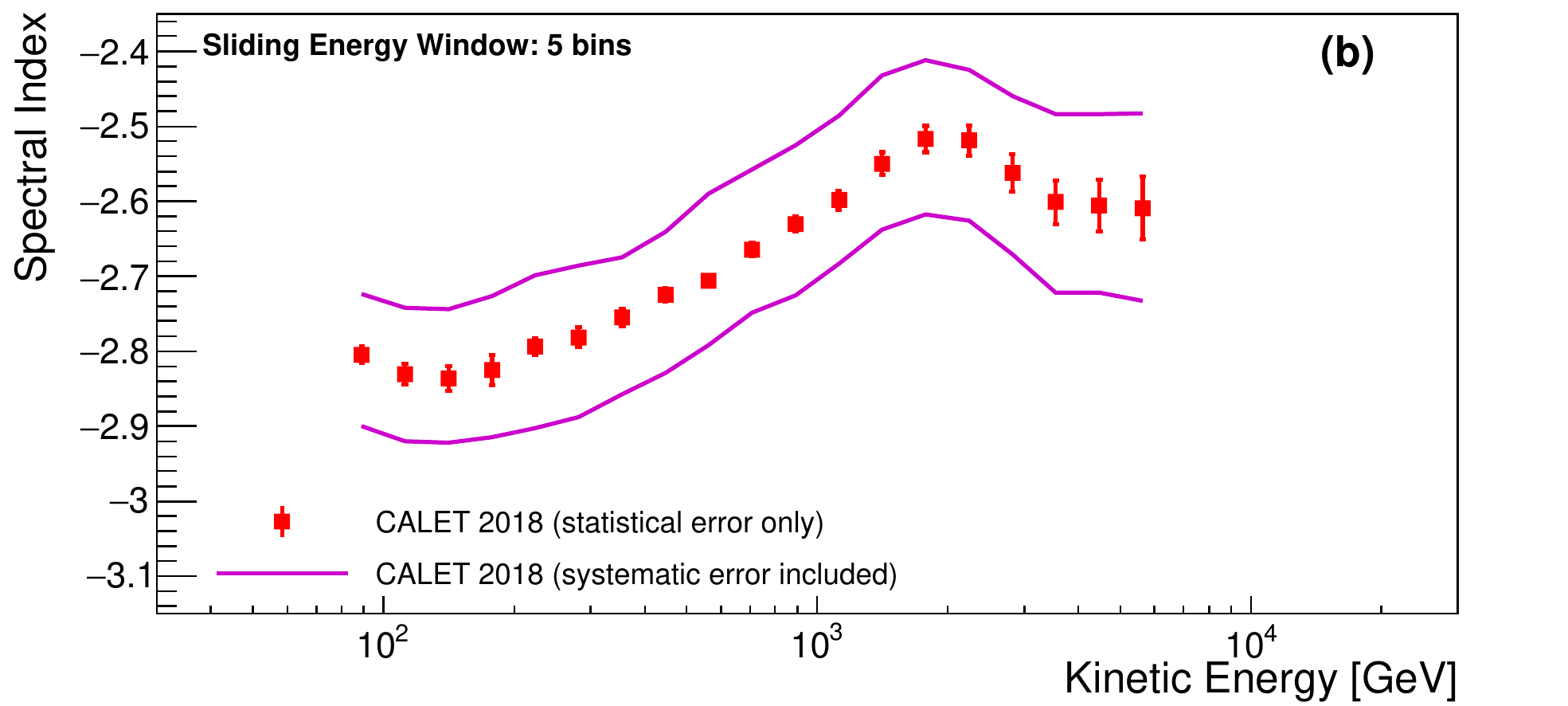}
\caption{(a) 
Fit 
of the CALET proton spectrum with 
single 
power-law 
functions. 
Red, blue, and magenta lines indicate the fit result for the energy ranges
between 50 and 500~GeV, 1 and 10~TeV, and 50~GeV and 10~TeV, respectively.
(b) Energy dependence of the spectral index calculated 
within 
a sliding energy window for CALET 
(red squares). 
The spectral index is determined for each bin by 
fitting the data 
using $\pm$2 energy bins. 
Magenta curves indicate the uncertainty range including systematic errors.
}
\label{fig:protonSpec}
\end{center}
\end{figure}

Figure~\ref{fig:protonSpec} (a) shows the 
fits 
of the CALET proton spectrum 
with a single power law. 
In order to study 
the 
spectral behavior, only the energy dependent systematics 
are included in the data points. 
Red, blue, and magenta lines indicate the fit result for the energy 
intervals 
between 50 and 500~GeV, 1 and 10~TeV, and 50~GeV and 10~TeV, respectively. 
The 
fit 
yields $\gamma_{1} = -2.81 \pm 0.03$ 
at 
lower energy (neglecting solar modulation effects) 
and 
$\gamma_{2} = -2.56 \pm 0.04$ 
at 
higher energy with good chi-square values.
On the other hand, the whole range fit gives 
a large 
chi-square per 
degree
of freedom, disfavoring the single 
power-law 
hypothesis 
by 
more than 3$\sigma$.
Our spectrum can also be fitted with 
a 
smoothly broken power-law function~\cite{AMS-02-proton,FF} as shown 
in 
Fig.~S7
of the Supplemental Material~\cite{CALET2018-proton-SM}, 
resulting in 
a power-law index of 
$-2.87 \pm 0.06$
(including solar modulation effects) below the breakpoint rigidity, which is
in good agreement with AMS-02~\cite{AMS-02-proton}. A larger variation of the 
power-law index of 
$0.30\pm0.08$
and a higher breakpoint rigidity of 
$496 \pm 175$~GV
than AMS-02~\cite{AMS-02-proton} are observed, though the latter 
is affected by relatively large error.

Furthermore, 
Fig.\ref{fig:protonSpec} (b) shows 
the energy 
dependence of the spectral index calculated 
within 
a sliding energy window (red squares).
The spectral index is determined for each bin by 
a fit of the data 
including the neighbor $\pm$2 bins. 
Magenta curves indicate the uncertainty 
band 
including systematic errors.
This result confirms a clear hardening of the spectrum above 
a few hundred GeV. 
These results may be important for the interpretation of 
the proton spectrum 
(e.g.~\cite{PR1-Blasi-2012, PR2-Aloisio-2013, Evoli2018}), 
since they indicate a 
progressive hardening up to the TeV region, 
while in good agreement with
magnet spectrometers in the 100 GeV to sub-TeV region.

\section{Conclusion}
We have measured, for the first time with an experimental apparatus in low Earth Orbit, 
the cosmic-ray proton spectrum from 50~GeV to 10~TeV,
covering with a single instrument the whole energy range previously investigated 
by magnetic spectrometers (BESS-TEV, PAMELA and AMS-02) 
and calorimetric instruments (ATIC, CREAM and NUCLEON) 
covering, in most of the cases, separate subranges of the region 
explored so far by CALET.
Our observations confirm the presence of a spectral hardening above a few hundred GeV. 
Our spectrum is not consistent with 
a 
single power law 
covering the whole range, 
while both
50--500~GeV and 1--10~TeV 
subranges 
can be 
separately 
fitted with single power-law functions, 
with the spectral index of the lower (higher) energy region 
being 
consistent 
with AMS-02~\cite{AMS-02-proton} (CREAM-III~\cite{CREAM-III-pHe}) within errors.
With the observation of a 
smoothly broken power law 
and of an energy dependence of the 
spectral index, CALET's proton spectrum will 
contribute to shed light on the origin of 
the spectral hardening.
Improved statistics and better understanding of the instrument
based on the analysis of additional flight data during the ongoing 
five years (or more) 
of observations 
might reveal 
a 
charge dependent energy cutoff 
possibly 
due to the acceleration limit
in supernova remnants in proton and helium 
spectra, 
or set
important constraints on the acceleration 
models. 

\section{Acknowledgments}
\begin{acknowledgments}
We gratefully acknowledge JAXA's contributions to the development of CALET and to the
operations onboard the International Space Station. 
We also 
express our sincere gratitude to ASI and NASA for
their support of the CALET project. This work was supported in part by JSPS Grant-in-Aid for Scientific Research (S) Grant No. 26220708, 
JSPS Grant-in-Aid for Scientific Research (B) Grant No. 17H02901, 
and by the MEXT-Supported Program for the Strategic
Research Foundation at Private Universities (2011--2015) (Grant No. S1101021) at Waseda University. 
The CALET effort in the United States is supported by NASA through Grants No. NNX16AB99G, No. NNX16AC02G, and No. NNH14ZDA001N-APRA-0075.
\end{acknowledgments}

\providecommand{\noopsort}[1]{}\providecommand{\singleletter}[1]{#1}  
\widetext
\clearpage
\begin{center}
\end{center}
\setcounter{equation}{0}
\setcounter{figure}{0}
\setcounter{table}{0}
\setcounter{page}{1}
\makeatletter
\renewcommand{\theequation}{S\arabic{equation}}
\renewcommand{\thefigure}{S\arabic{figure}}
\renewcommand{\bibnumfmt}[1]{[S#1]}
\renewcommand{\citenumfont}[1]{S#1}

\begin{center}
\textbf{\Large Direct Measurement of the Cosmic-Ray Proton Spectrum\\ 
 from 50~GeV to 10~TeV with the Calorimetric Electron Telescope\\
 on the International Space Station}

\end{center}
\vspace*{0.5cm}
Supplemental material 
relative to 
``Direct Measurement of the Cosmic-Ray Proton Spectrum
from 50~GeV to 10~TeV with the Calorimetric Electron Telescope
on the International Space Station.''
\vspace*{1cm}

\section{Data Analysis}
We describe the analysis procedure in three steps as follows.
Although some of the descriptions are duplicate with the main text, 
we have included them here for completeness.

\subsection{Event Selection}
The first step is selection of proton candidate events. 
The selection criteria to select proton events are optimized and
defined using MC simulations consisting of protons, helium and electrons.
The same criteria are applied to both of the Flight Data (FD) and MC data (MC). 

In order to minimize and to accurately separate protons from helium in charge identification,
it is important to preselect well reconstructed and well contained events. 
Furthermore, by removing events not included in the MC samples, i.e., 
those with incidence from zenith angle greater than 90$^\circ$ and mis-reconstructed events,
event samples equivalent between FD and MC were obtained to be fed into 
charge identification. 
This is the most important purpose of the preselection, 
which consists of (1) offline trigger confirmation, 
(2) geometrical condition,
(3) track quality cut,
(4) electron rejection cut,
(5) off-acceptance events rejection cut, 
(6) requirement of
track consistency with TASC energy deposits,
and (7) shower development requirement in IMC.
Each of the above selections are described in more detail in the following,
and finally the charge identification based on CHD and IMC energy deposits
is described. For the detailed description of the detector components used
in the event selections, readers are referred to the Supplemental Material
of Ref.~\cite{SM_CALET2017} and/or Refs.~\cite{SM_asaoka2017,SM_asaoka2018}.

\subsubsection{(1) Offline trigger confirmation}
A 
first event selection is the onboard high energy shower trigger (HE trigger). 
This trigger uses a simple trigger condition 
which selects showering particles above 10~GeV 
by requiring large energy deposits in the 
middle of the detector, i.e., energy deposit sums of 
IMC-X7$+$X8, IMC-Y7$+$Y8 and TASC-X1 to exceed certain thresholds 
in coincidence~\cite{SM_asaoka2018}.
Since the HE trigger is working onboard, 
it is affected by 
position dependence, temperature dependence, and temporal variation of the detector gain.
In order to obtain consistency between MC and FD in a simple way by 
removing such complicated effects, 
an offline trigger 
is applied as a first step of preselection, which requires sufficiently severer
conditions than the onboard HE trigger. 
After applying all the calibration, 
the offline trigger requires that the energy deposit sums of 
IMC-X7$+$X8, IMC-Y7$+$Y8 and TASC-X1 have to be greater than 50~MIP, 50~MIP and 100~MIP,
respectively, where one MIP corresponds to the energy deposit of minimum ionizing vertical 
muons at 2~GeV, i.e. 1.66~MeV for CHD paddle, 0.145~MeV for IMC fiber and 20.47~MeV 
for TASC log.

When analyzing events triggered by low-energy (LE) trigger 
which selects showering particles above 1~GeV, 
the offline trigger confirmation uses lower thresholds, i.e., 5~MIP for IMC-X7$+$X8, IMC-Y7$+$Y8 
and 10~MIP for TASC-X1.

\subsubsection{(2) Geometrical condition}
In order to ensure the 
accuracy of 
charge selection and energy measurement,
it is required that the 
reconstructed track must pass through the whole detector, 
i.e., from CHD top to TASC bottom, 
with 2~cm margin from the sides of the TASC, 
which is defined as Acceptance A.
All the geometrical conditions are summarized in the Supplemental Material of Ref.~\cite{SM_CALET2018}
for reference.

\subsubsection{(3) Track quality cut}
Combinatorial 
Kalman Filter 
(KF) 
tracking~\cite{ICRC2017_KF} was developed 
to reconstruct the proton tracks in a highly efficient way,
and is used in this analysis.
In order to ensure track quality, the algorithm for the tracking
is required to be KF tracking or shower fit in X-Z and Y-Z projection, 
and the 
$\chi^{2}$ 
of the 
fits to be less than 10 in 
both projections. 

\subsubsection{(4) Electron rejection cut}
In order to reject electrons especially in the lowest energy region,
Moliere concentration along the track is defined as follows:
for each IMC layer crossed by the track, 
a 
Moliere concentration is calculated summing all energy deposits found inside 
one 
Moliere radius 
($\pm$9 fibers) of each fiber matched to the track. Then the energy deposit sum within 
one 
Moliere radius
is divided by the total energy deposit sum in IMC.
By requiring this quantity to be less than 0.7, most of electrons are rejected while keeping very high
efficiency for protons.

\subsubsection{(5) Off-acceptance events rejection cut}
Off-acceptance events are defined as those reconstructed as Acceptance A,
but 
for which the true acceptance does not fulfill the condition of Acceptance A. 
Rejection of such off-acceptance events are necessary to 
precisely determine
geometrical
acceptance. The off-acceptance events consist dominantly of protons and helium. 
Since off-acceptance helium might not be separated in the 
charge selection due to the fact that secondaries (mostly pions) have
charge one, 
such helium contamination also needs to be minimized.

The off-acceptance cut uses two discrimination variables. 
The first variable is the maximum fractional energy deposit in a single TASC layer. 
It is 
required to be less than 0.4 to reject laterally incident events.
This selection is especially effective for TASC-X1 because it is used for trigger.
The second variable is the maximum energy deposit ratio of the edge 
logs 
to the maximum
log 
in each layer.
Events are rejected if this variable is greater than 0.4. This cut is effective to
remove events which exit from the side of TASC.
Those selections have very high efficiency while not depending on the track
reconstruction.

\subsubsection{(6) Requirement of track consistency with TASC energy deposits}
In order to further reject mis-reconstructed events, a consistency cut is defined between tracks and
centers of gravity of energy deposits in TASC-X1 and TASC-Y1 layers.
Energy dependent thresholds are defined using MC simulation 
to have a constant efficiency of 95\% for events 
that interacted in IMC below the 4th layer, which are suitable 
for determining charge, energy, and trigger efficiency (hereafter denoted as ``target'' events).

\subsubsection{(7) shower development requirement in IMC}
Since a fraction of events triggered by backscattering 
is not reproduced well by the simulations, 
rejection of such events is important.
For this purpose,
the energy deposit sum along the shower axis over $\pm$9 
fibers (in total 19 fibers) is 
used to ensure the existence of a 
shower core in IMC.
This definition differs from 
the one used for electrons considering the 
wider lateral spread of hadronic showers.
In order to fully exploit the rejection capability of events 
triggered by backscattering, 
it is important to set an appropriate threshold as a function of energy.
Energy dependent thresholds are defined to get 99\% efficiency for ``target'' events.

\subsubsection{Charge identification}
Based on the preselected samples, charge identification is performed
using the CHD and the IMC~\cite{SM_pier2017}. 
The latter samples the ionization deposits in each layer, thereby providing a
multiple d$E$/d$x$ measurement with a maximum of 16 samples along the track. 
The interaction point is first reconstructed~\cite{SM_brogi2015} 
and only the d$E$/d$x$ ionization clusters from the layers upstream 
the interaction point are used. The charge value is evaluated as 
a truncated-mean of the valid samples with a truncation level set at 70\%.

\begin{figure}[h!] 
\begin{center}
\includegraphics[width=\hsize]{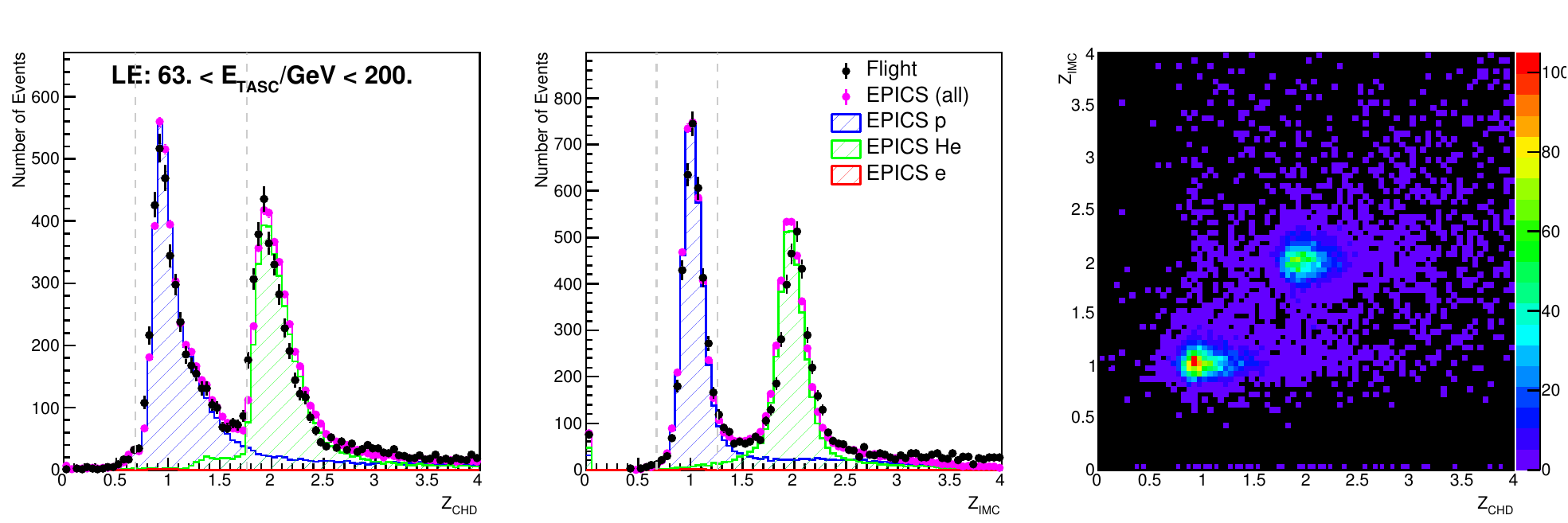}
\includegraphics[width=\hsize]{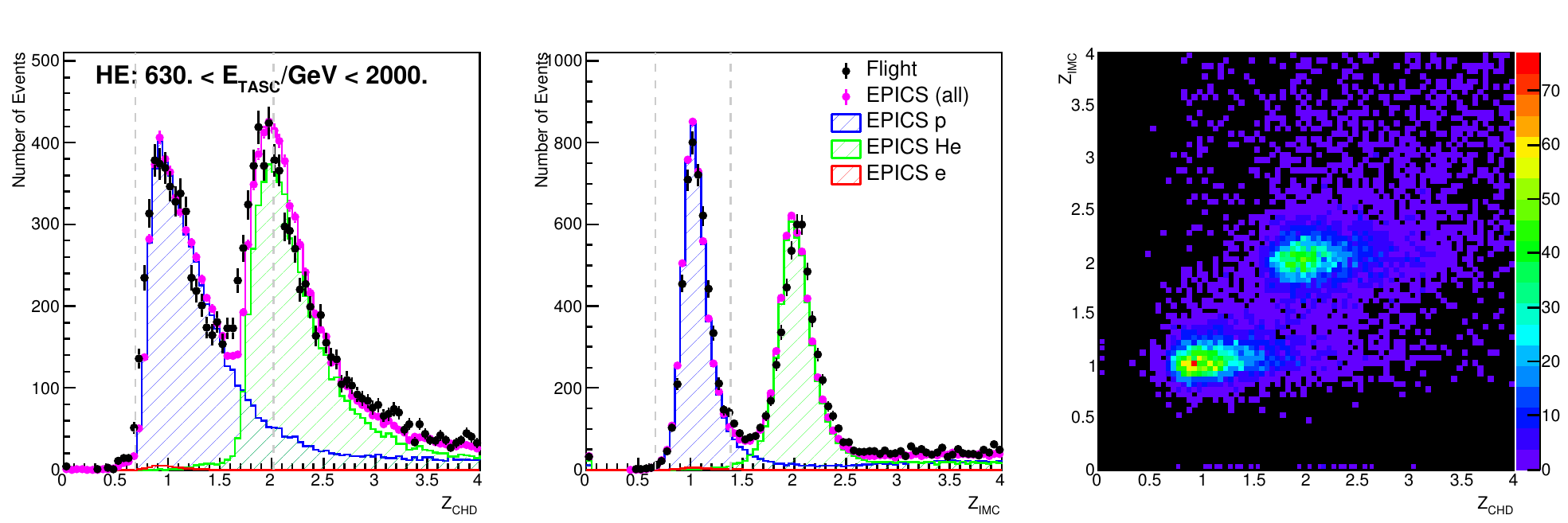}
\includegraphics[width=\hsize]{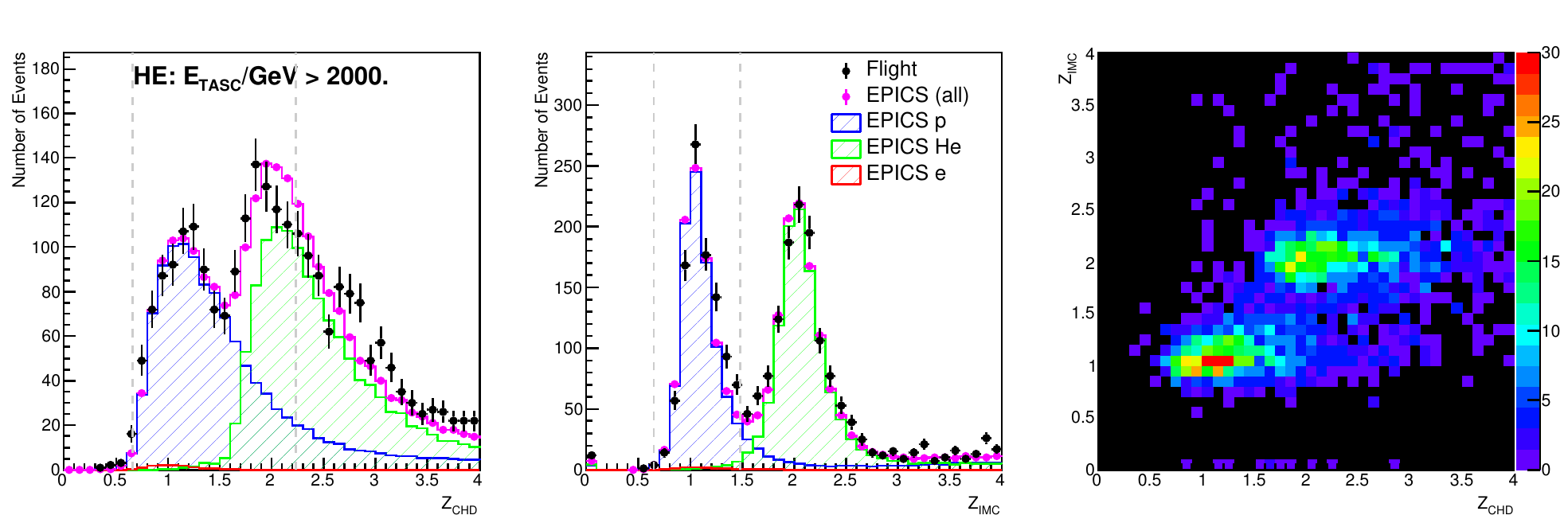}
\caption{Examples of CHD/IMC charge distributions.
Left, center and right panels show the CHD charge, IMC charge 
and correlation between CHD and IMC charges, respectively. 
From top to bottom, the plots 
corresponds to events with $63 < E_{\rm TASC} < 200$~GeV,
$630 < E_{\rm TASC} < 2000$~GeV, and $E_{\rm TASC} > 2000$~GeV, 
respectively. 
Gray dashed lines indicate the cut positions at 
110~GeV (Top), 1100~GeV (Middle) and 3600~GeV (Bottom).
}
\label{fig:distZ_SM}
\end{center}
\end{figure}

To mitigate the backscattering effects, an energy dependent charge correction to 
restore the nominal peak positions of protons and helium to $Z=1$ and 2 is applied 
separately to FD, EPICS, FLUKA and Geant4.
Charge selection of proton and helium candidates is performed 
by applying simultaneous window cuts on CHD and IMC reconstructed charges. 
The resultant charge distributions are 
exemplified in Fig.~\ref{fig:distZ_SM}.
For the selection with CHD and IMC, energy dependent thresholds 
are defined separately to keep 95\% efficiency for ``target'' events.

\subsection{Background Contamination}
Background contamination 
is  estimated from the MC simulation of protons, helium and electrons
as a function of observed energy, 
where the previous observations, i.e., 
AMS-02~\cite{SM_AMS-02-proton, SM_AMS-02-helium} and CREAM-III~\cite{SM_CREAM-III-pHe},
are used 
to simulate their spectral shape. 
Among them, the dominant component is off-acceptance protons except for the highest
energy region $E_{\rm TASC} \sim 10$~TeV, 
where helium contamination becomes dominant. 
Overall contamination is estimated 
below a few percent, and at maximum $\sim$5\% in the lowest and highest energy region.
The correction is carried out before performing 
the energy unfolding procedure.

In the lower energy region, the mis-reconstruction probability
for protons is higher, 
due to the poorer reconstruction of the TASC shower axis caused by the 
less prolate shower shapes at these energies. 
For helium, this mis-reconstruction probability is much
lower due to larger energy deposits in each hit produced by 
a 
primary 
track in IMC. This is the main reason behind the higher contamination
ratio due to off-acceptance protons in the low energy region.
In the higher energy region above 1 TeV, the effect of 
backscattering gets more and more significant and 
therefore 
the 
helium dominates the total contamination at the highest energy 
region although it is still sufficiently small not to significantly 
influence the proton spectrum.

\subsection{Energy Unfolding}
In order to take into account the relatively limited energy resolution,
energy unfolding is necessary to correct for bin-to-bin migration effects.
For reference, the observed energy fraction is around 35\% and 
the resultant energy resolution is 30--40\% in the energy region analyzed 
here.
As an energy unfolding method in this analysis, we used 
the Bayesian approach implemented in the RooUnfold 
package~\cite{SM_roounfold, SM_BayesUnfolding} 
in ROOT~\cite{SM_root}, 
with the response matrix derived using MC simulation. 
Convergence is obtained within two iterations, 
given the relatively accurate 
prior distribution obtained from 
the previous observations, i.e., AMS-02~\cite{SM_AMS-02-proton} and 
CREAM-III~\cite{SM_CREAM-III-pHe}.

Though CALET calorimeter is homogeneous, practically most of 
calorimeters are non-compensating to a certain degree. 
Therefore, a correction for electrons is not necessarily 
the same as for protons.
Because of the limited energy resolution, an absolute energy 
scale calibration using geomagnetic rigidity cutoff used
in Refs.~\cite{SM_CALET2017, SM_CALET2018} could not be performed.

\subsection{Consistency between LE and HE Analyses}
\begin{figure}[h!]
 \centering
 \begin{minipage}{1.0\hsize}
 \begin{center}
  \includegraphics[width=\hsize]{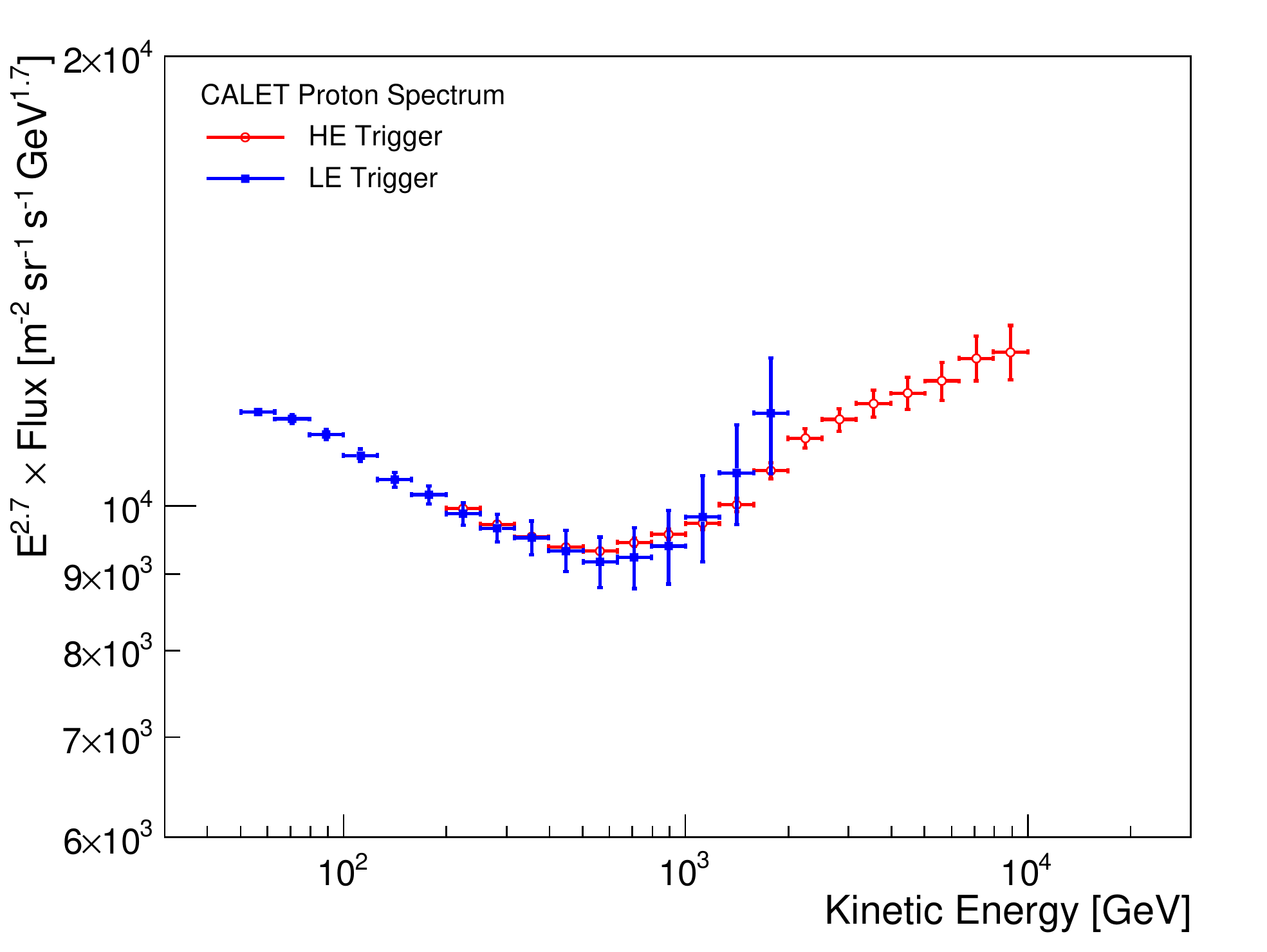}
  \caption{
Proton spectrum from two data sets 
corresponding 
to LE-trigger (blue squares) and HE-trigger (red circles) analyses.}
  \label{fig:flxcmp}
 \end{center}
 \end{minipage}
\end{figure}

Depending on the on-orbit trigger mode and corresponding offline-trigger threshold,
two spectra are obtained with the LE and HE analyses, respectively, 
as shown in Fig.~\ref{fig:flxcmp}.
For $E < 200$~GeV, the use of LE-trigger analysis is required because 
an offline trigger threshold higher than in the hardware trigger
was found to introduce an efficiency bias in the HE-trigger analysis, 
which became evident with a scan of the offline-trigger threshold 
using LE-trigger data.
Since both fluxes are well consistent in $E > 200$~GeV, 
they are combined around $E \sim 300$~GeV, 
taking into account the different statistics of the two trigger modes. 

\section{Systematic Uncertainties}
Dominant sources of systematics uncertainties in proton analysis include: 
\begin{itemize}
\item[(1)] hadronic interaction modeling, 
\item[(2)] energy response,
\item[(3)] track reconstruction, and 
\item[(4)] charge identification.
\end{itemize}
To address these uncertainties, various approaches are used and discussed 
in the following. 
An 
important part of systematics comes from the accuracy
of the beam test calibration and its extrapolation/interpolation.
The 
stability of the measured spectrum against variations of several analysis cuts 
is also a
crucial tool to estimate the associated 
uncertainties. 

Most of the systematic uncertainties in normalization 
are taken from the studies performed for the electron analysis. 
This uncertainty is 
estimated 
as 4.1\% based on the electron spectrum paper~\cite{SM_CALET2017}, 
as the sum in quadrature of the uncertainties on 
live time (3.4\%), radiation environment (1.8\%), and
long-term stability (1.4\%).

\noindent {\bf Hadronic interaction: }
The uncertainty in the hadronic interaction affects directly the 
trigger efficiency and it is also closely connected to the uncertainty 
in the energy response, 
as 
discussed separately in the following. 
In the low-energy region, the absolute calibration of the trigger efficiency 
was performed 
at 
the beam test. The main source of uncertainty comes from the 
accuracy of the calibration. 
In addition to the measurement accuracy, possible systematic bias 
due to normalization in the measurements of trigger efficiency 
was considered as 
a systematic uncertainty 
and is estimated as 2.2\% and 3.3\% for HE and LE analyses, 
respectively. 

\begin{figure}[hbtp]
 \centering
 \begin{minipage}{1.0\hsize}
 \begin{center}
  \includegraphics[width=\hsize]{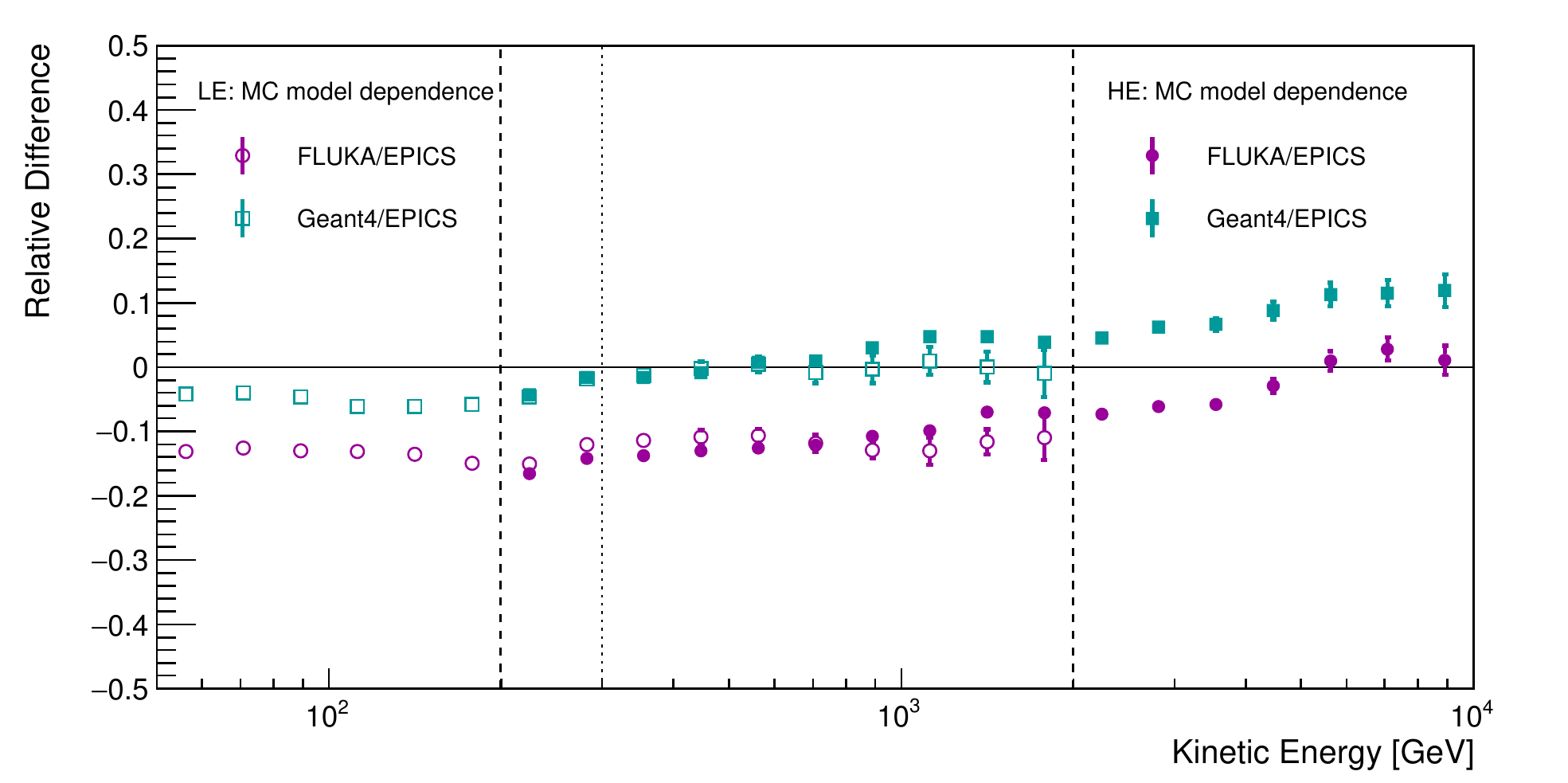}
  \caption{
Relative difference of
    the measured 
    flux 
when different MC simulations are used
with respect to the EPICS case.
    FLUKA/EPICS and Geant4/EPICS are shown as magenta circles and cyan squares, respectively.
    Open (closed) symbols are used for LE-trigger (HE-trigger) analysis.
The upper (lower) bound of the energy region used in LE 
(HE) analysis is shown by the dashed line on the left (right) side 
of the picture. 
The two analyses are combined at 300~GeV as indicated
by dotted line.
The relatively larger difference between FLUKA and EPICS 
appears to be caused by different normalization as FLUKA and Geant4 
curves are practically parallel. 
Considering the fact that EPICS has been validated 
with beam-test measurements 
in the low-energy region, we have conservatively
attributed one-half of the difference between FLUKA/Geant4 and 
EPICS as energy dependent systematic uncertainty.
}
  \label{fig:sysMC}
 \end{center}
 \end{minipage}
 \centering
 \begin{minipage}{1.0\hsize}
 \begin{center}
  \includegraphics[width=\hsize]{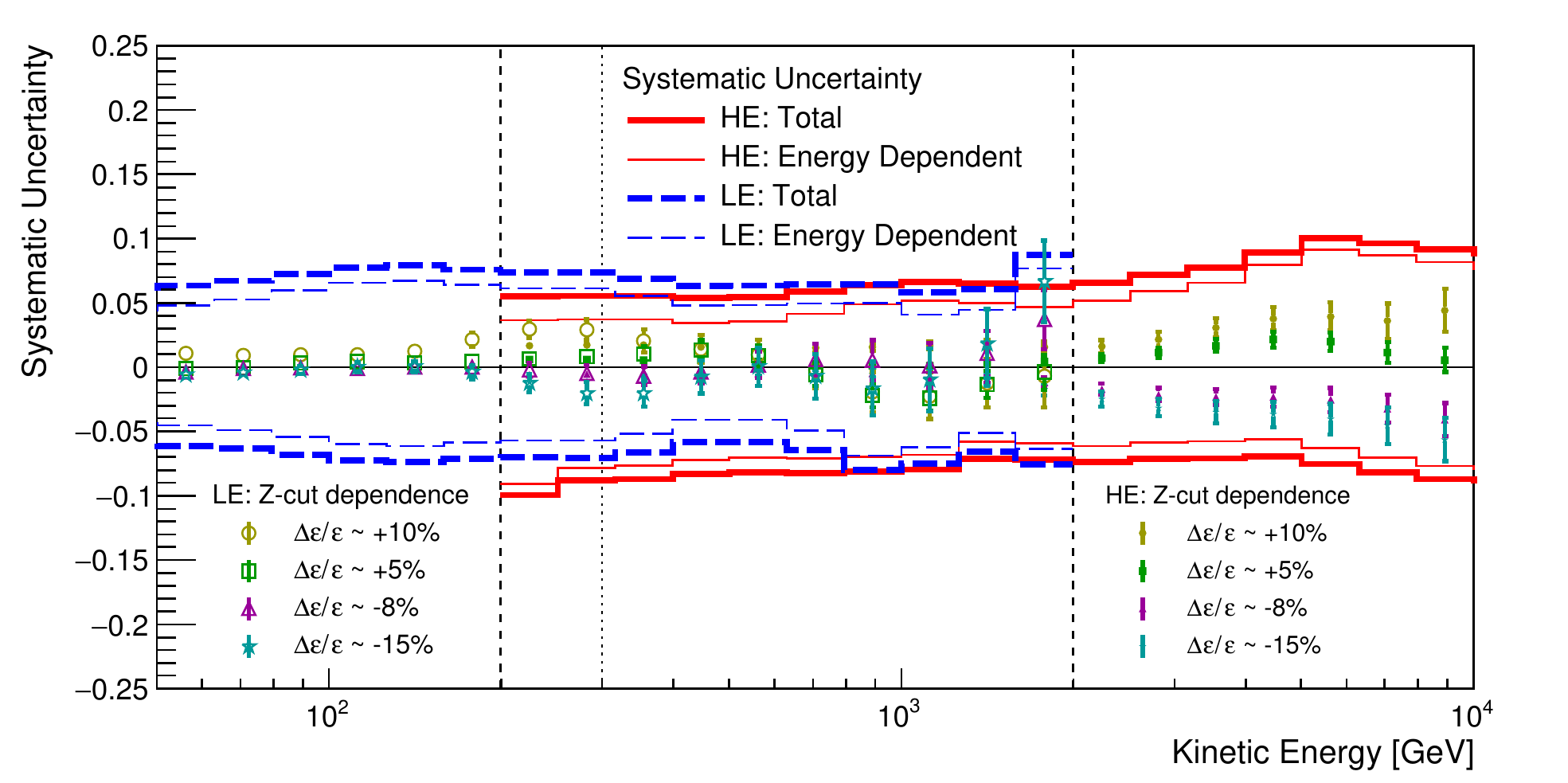}
  \caption{
    Energy dependence of total systematic uncertainty band (thick lines).
    Contributions from energy dependent systematic uncertainty 
    are 
    shown as 
thinner lines of the same color.
    Blue long-dashed 
    lines and red solid lines 
    correspond 
    to LE- and HE-trigger analyses, respectively.
Relative differences of the measured flux in the case of
different charge selections with respect to the nominal charge cut
(the efficiency variation considered here ranges from 
$\sim -15$\% to $\sim +10$\%) 
are also shown. The standard
deviation of the relative differences is taken as energy dependent 
systematic uncertainty in each energy bin. The total uncertainty is 
the quadratic sum of energy dependent and normalization uncertainties.
The upper (lower) bound of the energy region used in LE 
(HE) analysis is shown by the dashed line on the left (right) side 
of the picture. 
The two analyses are combined at 300~GeV as indicated
by dotted line.
  }
  \label{fig:sysTotal}
 \end{center}
 \end{minipage}
\end{figure}

In the high-energy region, 
a non-trivial 
extrapolation from the maximum 
available
beam energy,
i.e. 400~GeV, is necessary. To address this uncertainty, the relative 
differences 
between different MC models, i.e., 
FLUKA and Geant4 versus EPICS
were 
investigated as shown
in Fig.~S\ref{fig:sysMC}. 
In the FLUKA and Geant4 simulation, the same corrections for
EPICS 
were applied, as determined from the beam test data. 
It should be noted that other effects such as the difference in backscattering 
treatment and energy responses are also 
included 
in 
this study.
Considering 
that (i) there 
is 
good consistency between
LE and HE analyses, (ii) EPICS is directly calibrated with 
beam test data, 
(iii)
backscattering at higher energies 
is
better simulated in EPICS than 
in 
Geant4 and FLUKA~\cite{SM_CALET2017} 
and (iv) the difference 
in 
the energy response among 
the 3 MC models show only little energy dependence
one half of the differences have been included 
in the energy dependent systematics
(see the comments in the caption of Fig.~S\ref{fig:sysMC}). 
The difference between DPMJET-III (reference) and EPOS 
was 
also studied, but
it 
was found to be 
completely negligible in the energy range considered here, mainly
because the use of EPOS 
is 
allowed 
only 
above 20~TeV.

\noindent {\bf Energy response: }
The uncertainty in the energy response is closely related to the 
uncertainty in the 
modeling of the hadronic 
interaction.
As in the case of the uncertainty of 
the trigger efficiency, 
the absolute calibration of the energy response
was performed 
using the beam test data 
in the low-energy region.
The main source of uncertainty in the energy response comes from the 
accuracy of the calibration,
with dominant 
contributions from the uncertainty in temperature of $\pm$0.5$^o$,
which %
translates 
into 2.8\% energy scale uncertainty.

As in the beam test analysis only 3 TASC logs per layer 
were 
used, 
the difference 
of the spectrum obtained with energy measurements 
between 3 TASC logs (the 
one 
associated with the track 
and the 
two lateral 
neighbors) 
and the whole TASC sum (used in this analysis)
is considered as the correction factor. 
In addition to that, 
one half 
of the 
correction
is included as upper and 
lower systematic error.

In the high-energy region, significant extrapolation from the maximum beam energy,
i.e. 400~GeV, is necessary,
which is 
taken into account as 
MC model dependence.

While the beam test correction basically 
addresses 
the relation between the primary energy and the 
mean shower energy, the effect of energy resolution should also be considered.
Separate unfolding procedures 
with TASC energy sum 
including the 
log
being hit $\pm$2, $\pm$3, and $\pm$5 neighbors are 
applied and the stability of the 
spectrum 
is 
included in the systematics, where stability is defined as the standard 
deviation of the relative differences 
in each energy bin 
with respect to the 
reference 
flux. 

\noindent {\bf Track reconstruction: }
It is 
not easy 
to directly 
assess 
the uncertainty in
track reconstruction. 
However, since tracking is the basis of 
most of the analysis, 
especially for the track-dependent selection cuts,
the effects 
are evaluated by studying the dependence 
on the charge cut and 
some pre-selection cuts, 
especially (2), 
(6) and (7).
To investigate the uncertainty in 
the definition of the acceptance, 
restricted acceptance regions 
are 
studied 
and the resultant fluxes
are compared, 
resulting in
negligible differences.
Regarding cut (6), 
efficiencies 
were 
varied by $\sim +20$\%, 
$\sim +10$\%, $\pm 0$\%, $\sim -20$\%, and $\sim -40$\% 
(corresponding 
to 99\%, 97\%, 95\%, 90\% and 85\% 
efficiencies for "target" events), and the relative differences 
with respect to the reference cases 
were 
obtained for each energy bin.
The standard deviation of the relative differences 
were 
considered 
as systematic uncertainty associated with cut (6). 
As per cut (7), 
a 
tighter cut is used 
with an 
efficiency for "target" events 
of 
95\% instead of 99\%. The relative differences with 
respect to the nominal case (99\% efficiency) are considered as the 
systematic 
uncertainty, 
which are applied to both positive and 
negative sides.

\noindent {\bf Charge identification: }
As 
helium contamination is one of the main uncertainties in the proton spectrum 
analysis especially in the high energy region, it is very important to 
study 
the
flux 
stability against 
charge cut efficiencies 
considering that the 
contamination ratio from
helium 
may depend on the same cuts. 
The stability against the charge cut efficiency is shown 
in Fig.~S\ref{fig:sysTotal}
for LE- and HE-trigger analyses.
They are included in the systematic uncertainty.

\begin{figure}[hbtp]
 \centering
 \begin{minipage}{1.0\hsize}
 \begin{center}
  \includegraphics[width=\hsize]{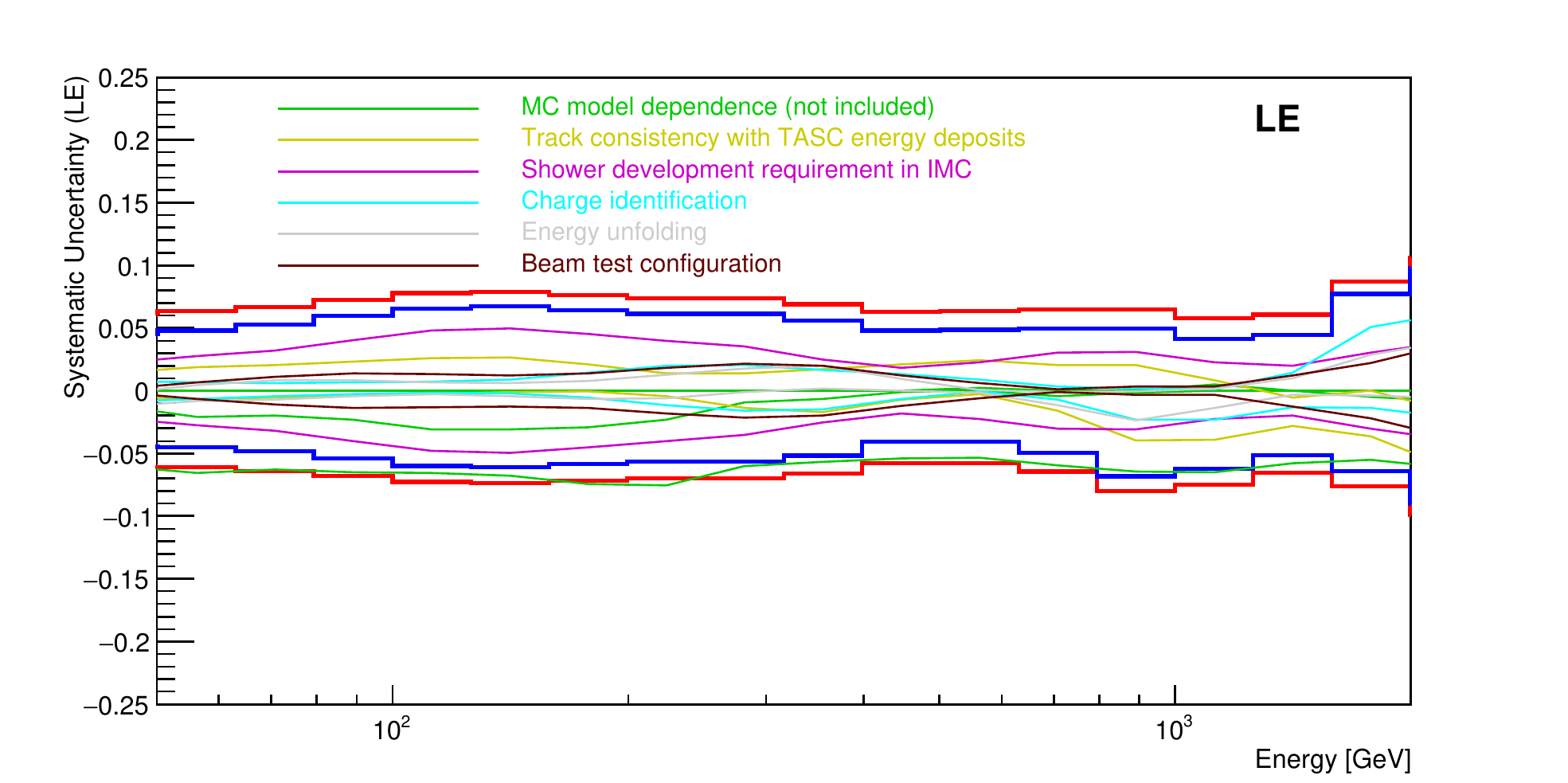}
  \includegraphics[width=\hsize]{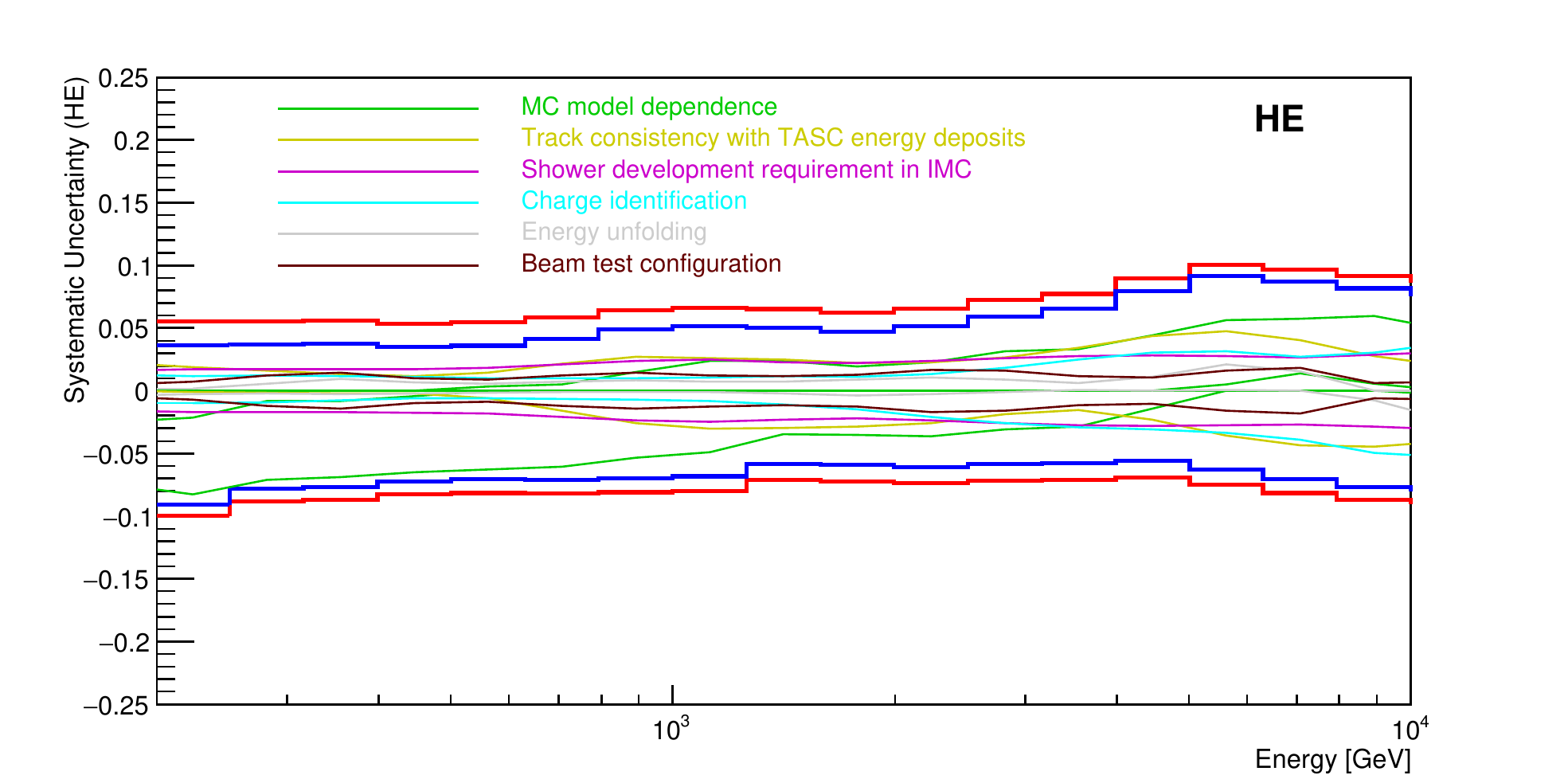}
  \caption{
    Summary of energy dependent systematic uncertainties
    for LE-trigger 
(upper figure) 
and HE-trigger 
(lower figure) 
analysis,
    where 
the 
energy dependence of total systematic uncertainty bands are 
shown 
by the thick red lines, while the thick blue lines isolate the total
contribution from systematic uncertainties with an energy dependence. 
    Thin lines with other colors show the contributions from each component
as explained in the legend. 
The energy 
dependent systematic uncertainty includes the contributions 
    from trigger efficiency which is estimated as 3.3\% and 2.2\% for LE and HE
    analyses, respectively.
}
  \label{fig:sysEDep}
 \end{center}
 \end{minipage}
\end{figure}

\noindent {\bf Total systematic uncertainty: }
Considering all of the above contributions, blue long-dashed and red solid lines
of Fig.~S\ref{fig:sysTotal} show the total systematic
uncertainty for LE and HE analyses, respectively, 
as a function of primary energy in the proton spectrum analysis.
For reference, a breakdown of the individual energy dependent 
systematic uncertainties in LE and HE analyses 
is shown in Fig.~S\ref{fig:sysEDep}.

\clearpage
\section{Results}
\begin{figure*}[bth!]
\begin{center}
\includegraphics[width=\hsize]{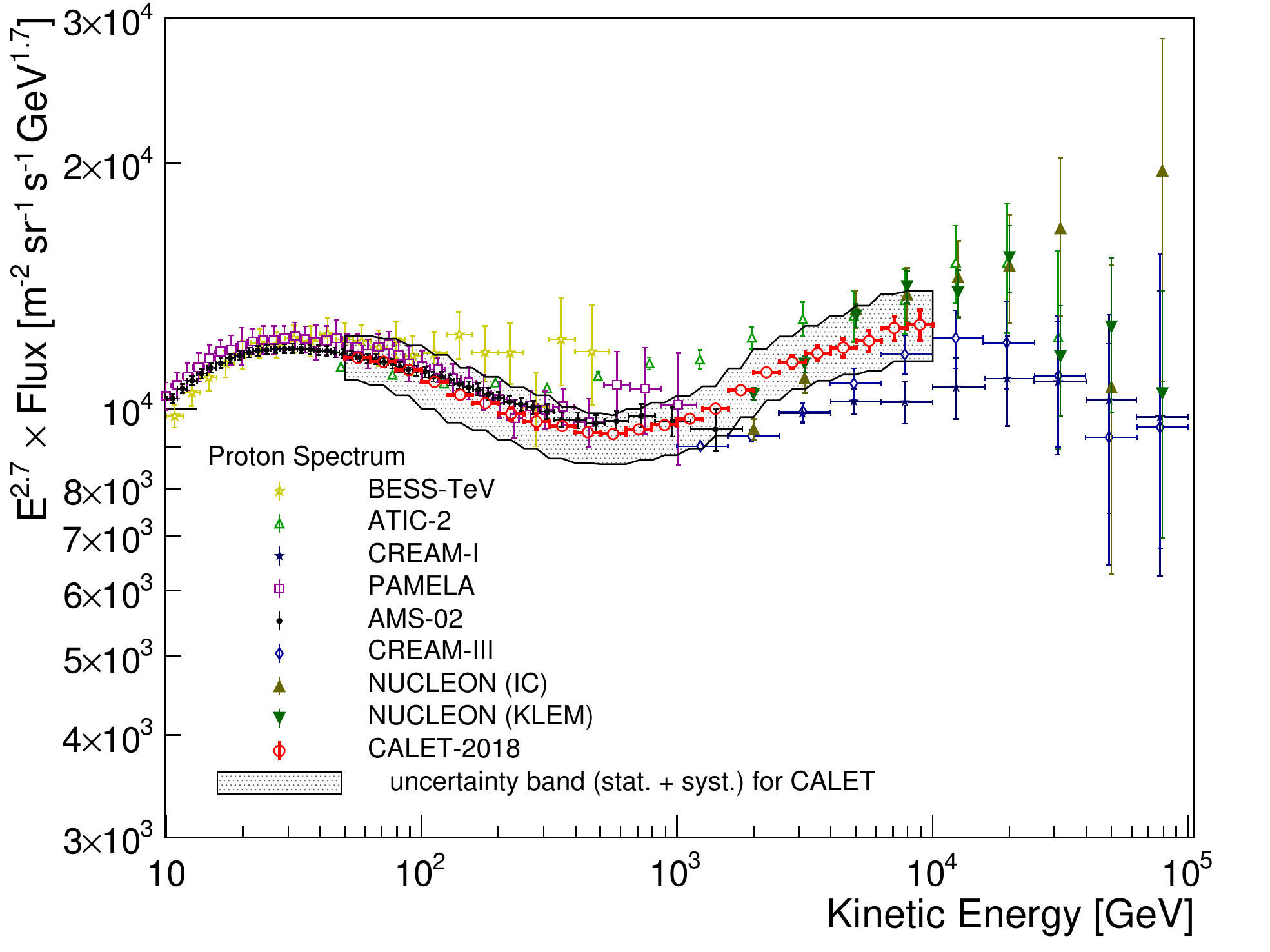}
\caption{Cosmic-ray proton spectrum measured by CALET from 50~GeV to 10~TeV 
using 
an 
energy binning of 10~bins per 
decade. 
The 
gray band indicates the quadratic sum of statistical and systematic errors. 
Also plotted are recent direct measurements in 
space~\cite{AMS02-p,SM_PAMELA-rep,SM_PAMELA-10yrs,SM_NUCLEON-JTEP} 
and from 
high altitude balloons~\cite{SM_BESS-TeV,SM_ATIC2-p,CREAM-I,SM_CREAM-III-pHe}.  
} 
\label{fig:protonSM}
\end{center}
\end{figure*}

\begin{figure}[htb!]
\begin{center}
\includegraphics[width=\hsize]{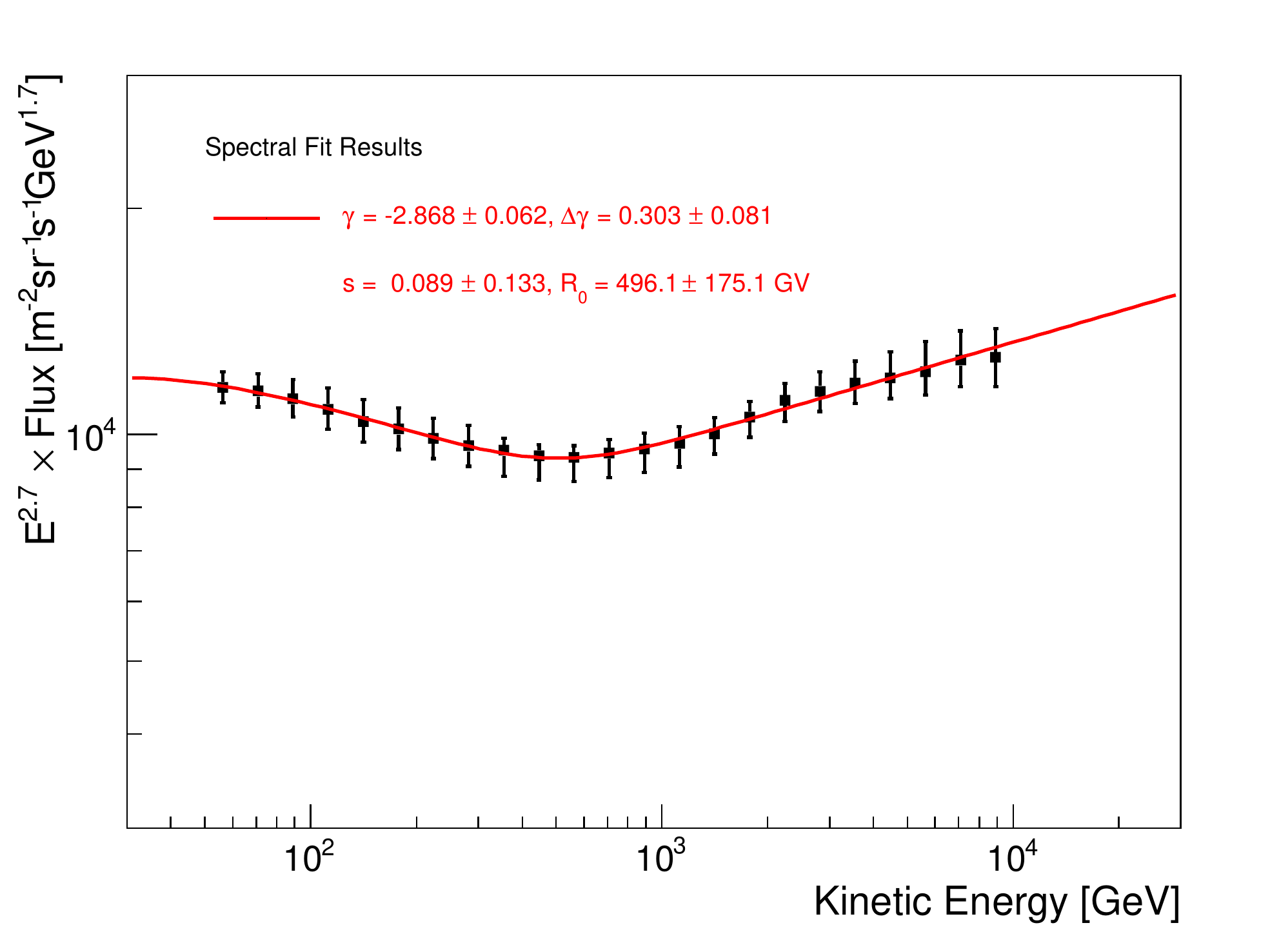}
\caption{
Fit 
of the CALET proton spectrum with a smoothly broken power-law function
as defined in Eq. (3) of Ref.~\cite{AMS02-p}.
Statistical errors are shown in quadrature with 
systematic errors including only energy dependent ones, neglecting 
to first order possible correlations among different sources of 
systematic errors and assuming that they are normally distributed.
The fit parameters are given in the plot. 
Since the fit 
is performed as a function of rigidity, an appropriate 
change of variables is carried out in the fitting function. 
The softer value of the spectral index obtained in this 
smooth 
fit 
with respect to the single power-law fit 
in the 50-500 GeV energy interval
is mainly 
due to 
the
inclusion of 
a
solar modulation potential of 
550~MV 
(fixed) 
using the force field approximation~\cite{SM_FF}. 
The stability of the 
result versus the choice of the potential has been studied by varying 
its value from 450~MV to 650~MV.  The result has been found to be 
insensitive within errors, differing at most by 0.01 for 
$\gamma$ and $\Delta \gamma$.
}
\label{fig:one}
\end{center}
\end{figure}

\clearpage

\renewcommand{\arraystretch}{1.25}
\begin{table*}[hbtp!]
\caption{
Table of CALET proton spectrum. 
Energy in each bin is simply represented by 
geometric mean. 
The first and second errors associated with mean energy represent 
systematic error in global energy scale and potentially energy dependent
ones, respectively. Energy dependent energy scale error mainly comes 
from beam test in the lower energy region and energy calibration of 
lower gain range~\cite{SM_asaoka2017} in the higher energy region.
For the flux the first, second and third
errors represent the statistical uncertainties (68\% confidence level),
systematic uncertainties in normalization, and energy dependent 
systematic uncertainties, respectively.
\label{tab:pro}}
\begin{ruledtabular}
\begin{tabular}{cccrrrrrrr}
Energy Bin & Representative Energy & Flux \\
(GeV)  & (GeV) & (m$^{-2}$sr$^{-1}$s$^{-1}$GeV$^{-1}$) \\ 
\hline
50.1--63.1 & $56.2\pm1.6\pm0.6$ &$(2.176 \, \pm 0.012 \, \pm 0.089 \, _{-0.099}^{+0.105}) \times 10^{-1}$\\
63.1--79.4 & $70.8\pm2.0\pm0.7$ &$(1.157 \, \pm 0.008 \, \pm 0.048 \, _{-0.056}^{+0.061}) \times 10^{-1}$\\
79.4--100.0 & $89.1\pm2.5\pm0.9$ &$(6.064 \, \pm 0.047 \, \pm 0.249 \, _{-0.329}^{+0.361}) \times 10^{-2}$\\
100.0--125.9 & $112.2\pm3.1\pm1.1$ &$(3.153 \, \pm 0.030 \, \pm 0.130 \, _{-0.189}^{+0.207}) \times 10^{-2}$\\
125.9--158.5 & $141.3\pm4.0\pm1.4$ &$(1.631 \, \pm 0.019 \, \pm 0.067 \, _{-0.100}^{+0.110}) \times 10^{-2}$\\
158.5--199.5 & $177.8\pm5.0\pm1.8$ &$(8.558 \, \pm 0.122 \, \pm 0.352 \, _{-0.501}^{+0.546}) \times 10^{-3}$\\
199.5--251.2 & $223.9\pm6.3\pm2.3$ &$(4.463 \, \pm 0.078 \, \pm 0.183 \, _{-0.253}^{+0.274}) \times 10^{-3}$\\
251.2--316.2 & $281.8\pm7.9\pm2.9$ &$(2.345 \, \pm 0.050 \, \pm 0.096 \, _{-0.133}^{+0.143}) \times 10^{-3}$\\
316.2--398.1 & $354.8\pm9.9\pm3.6$ &$(1.242 \, \pm 0.004 \, \pm 0.051 \, _{-0.095}^{+0.046}) \times 10^{-3}$\\
398.1--501.2 & $446.7\pm12.5\pm4.5$ &$(6.565 \, \pm 0.029 \, \pm 0.270 \, _{-0.472}^{+0.228}) \times 10^{-4}$\\
501.2--631.0 & $562.3\pm15.7\pm5.8$ &$(3.505 \, \pm 0.018 \, \pm 0.144 \, _{-0.247}^{+0.125}) \times 10^{-4}$\\
631.0--794.3 & $707.9\pm19.8\pm7.5$ &$(1.907 \, \pm 0.012 \, \pm 0.078 \, _{-0.136}^{+0.079}) \times 10^{-4}$\\
794.3--1000.0 & $891.3\pm25.0\pm10.2$ &$(1.037 \, \pm 0.008 \, \pm 0.043 \, _{-0.072}^{+0.051}) \times 10^{-4}$\\
1000.0--1258.9 & $1122.0\pm31.4\pm14.5$ &$(5.665 \, \pm 0.050 \, \pm 0.233 \, _{-0.387}^{+0.292}) \times 10^{-5}$\\
1258.9--1584.9 & $1412.5\pm39.6\pm20.4$ &$(3.131 \, \pm 0.033 \, \pm 0.129 \, _{-0.182}^{+0.157}) \times 10^{-5}$\\
1584.9--1995.3 & $1778.3\pm49.8\pm27.8$ &$(1.772 \, \pm 0.023 \, \pm 0.073 \, _{-0.105}^{+0.083}) \times 10^{-5}$\\
1995.3--2511.9 & $2238.7\pm62.7\pm36.9$ &$(1.001 \, \pm 0.015 \, \pm 0.041 \, _{-0.061}^{+0.051}) \times 10^{-5}$\\
2511.9--3162.3 & $2818.4\pm78.9\pm48.4$ &$(5.531 \, \pm 0.095 \, \pm 0.227 \, _{-0.324}^{+0.328}) \times 10^{-6}$\\
3162.3--3981.1 & $3548.1\pm99.3\pm62.9$ &$(3.044 \, \pm 0.063 \, \pm 0.125 \, _{-0.176}^{+0.200}) \times 10^{-6}$\\
3981.1--5011.9 & $4466.8\pm125.1\pm81.4$ &$(1.661 \, \pm 0.041 \, \pm 0.068 \, _{-0.093}^{+0.132}) \times 10^{-6}$\\
5011.9--6309.6 & $5623.4\pm157.5\pm104.8$ &$(9.090 \, \pm 0.266 \, \pm 0.374 \, _{-0.571}^{+0.832}) \times 10^{-7}$\\
6309.6--7943.3 & $7079.5\pm198.2\pm134.9$ &$(5.056 \, \pm 0.174 \, \pm 0.208 \, _{-0.356}^{+0.441}) \times 10^{-7}$\\
7943.3--10000.0 & $8912.5\pm249.6\pm172.3$ &$(2.741 \, \pm 0.115 \, \pm 0.113 \, _{-0.210}^{+0.224}) \times 10^{-7}$\\
\end{tabular}
\end{ruledtabular}
\end{table*}
\renewcommand{\arraystretch}{1.0}

\providecommand{\noopsort}[1]{}\providecommand{\singleletter}[1]{#1}  

\begin{thebibliography}{61}
\bibitem [{EPI()}]{EPICSver}   \BibitemOpen  \href@noop {} {\bibinfo  {journal} {EPICS and COSMOS versions are 9.20 and  8.00, respectively}\ }\BibitemShut {NoStop}
\bibitem [{Gea()}]{Geant4ver}   \BibitemOpen  \href@noop {} {\bibinfo  {journal} {Geant4 version employs FTFP\_BERT as a  physics list which is recommended for the simulation of high energy showers}\  }\BibitemShut {NoStop}
\end{thebibliography}
\end{document}